\documentclass[aps,prx,preprint,groupedaddress,nofootinbib,reprint]{revtex4-2}

\usepackage{natbib}
\usepackage[utf8]{inputenc}
\usepackage[american]{babel}
\usepackage[T1]{fontenc}
\usepackage{amsmath,amssymb,amsfonts}

\usepackage{standalone}
\standaloneconfig{mode=image}
\usepackage{tikz}
\usetikzlibrary{positioning,arrows,calc,math,angles,quotes}
\tikzstyle{arrow} = [thick,->,>=stealth] 

\usepackage{amsmath}

\usepackage{hyperref}
\hypersetup{colorlinks=true, 
breaklinks=true, 
urlcolor= black, 
linkcolor= black,
citecolor= black 
} 

\usepackage{cleveref} 
\crefname{figure}{fig.}{fig.} 
\Crefname{figure}{Fig.}{Fig.} 
\crefname{equation}{}{}
\Crefname{equation}{}{}

\usepackage[colorinlistoftodos, textsize=small]{todonotes}

\newcommand{\ii}{{{\mathbf{i}}\mkern1mu}} 
\usepackage{braket} 
\newcommand\ketbra[1]{\ensuremath{ \ket{#1} \bra{#1} }}

\newcommand{\hc}{\ensuremath{\mathit{h.c.}}}
\newcommand{\qop}[1]{\widehat{#1}}
\newcommand{\binaire}[1]{\mathrm{b}[{#1}]}
\newcommand{\modulo}[1]{\mathrm{mod}[{#1}]}
\newcommand{\taille}[1]{\mathrm{size}[{#1}]}

\newcommand{\aide}[1]{\ensuremath{\scriptstyle \boldsymbol{\mathbf{#1}}}}

\usepackage{acro} 

\DeclareAcronym{sth}{short=\textsc{STH}, long=Special Tripotent Hamiltonian}
\DeclareAcronym{scb}{short=\textsc{SCB}, long=Single Component Basis}
\DeclareAcronym{ps}{short=\textsc{PS}, long=Pauli-String}
\DeclareAcronym{lcu}{short=\textsc{LCU}, long=Linear Combination of Unitaries}
\DeclareAcronym{lch}{short=\textsc{LCH}, long=Linear Combination of Hamiltonian}
\DeclareAcronym{pvm}{short=\textsc{PVM}, long=Projective Valued Measurement}
\DeclareAcronym{povm}{short=\textsc{POVM}, long=Positive Operator-Valued Measure}
\DeclareAcronym{sic}{short=\textsc{SIC}, long=Symmetric Informationally Complete}
\DeclareAcronym{csco}{short=\textsc{CSCO}, long=Complete Set of Commuting Observables}
\DeclareAcronym{hs}{short=\textsc{HS}, long=Hamiltonian Simulation}
\DeclareAcronym{be}{short=\textsc{BE}, long=Block-Encode}
\DeclareAcronym{bei}{short=\textsc{BEing}, long=Block-Encoding}
\DeclareAcronym{bed}{short=\textsc{BEed}, long=Block-Encoded}

\DeclareAcronym{qa}{short=\textsc{QA}, long=Quantum Arithmetic}
\DeclareAcronym{qft}{short=\textsc{QFT}, long=Quantum Fourier Transform}
\DeclareAcronym{iqpe}{short=\textsc{iQPE}, long=Iterative Quantum Phase Estimation}
\DeclareAcronym{qpe}{short=\textsc{QPE}, long=Quantum Phase Estimation}
\DeclareAcronym{hhl}{short=\textsc{HHL}, long=Harrow–Hassidim–Lloyd algorithm}
\DeclareAcronym{qsp}{short=\textsc{QSP}, long=Quantum Signal Processing}
\DeclareAcronym{qsvt}{short=\textsc{QSVT}, long=Quantum Singular Value Transformation}
\DeclareAcronym{qet}{short=\textsc{QET}, long=Quantum Eigenvalue Transformation}
\DeclareAcronym{gqsp}{short=\textsc{GQSP}, long=Generalized Quantum Signal Processing}
\DeclareAcronym{seqsp}{short=\textsc{SE-QSP}, long=Semantically Embedded Quantum Signal Processing}
\DeclareAcronym{qpp}{short=\textsc{QPP}, long=Quantum Phase Processing}
\DeclareAcronym{mqsp}{short=\textsc{M-QSP}, long=Multivariable Quantum Signal Processing}
\DeclareAcronym{vqa}{short=\textsc{VQA}, long=Variationnal Quantum Algorithm}
\DeclareAcronym{vqe}{short=\textsc{VQE}, long=Variationnal Quantum Eigensolver}
\DeclareAcronym{vqls}{short=\textsc{VQLS}, long=Variational Quantum Linear Solver}
\DeclareAcronym{ucc}{short=\textsc{UCC}, long=Unitary Coupled Cluster}
\DeclareAcronym{hva}{short=\textsc{HVA}, long=Hamiltonian Variational Ansatz}
\DeclareAcronym{qaoa}{short=\textsc{QAOA}, long=Quantum Approximate Optimization Algorithm}
\DeclareAcronym{aqc}{short=\textsc{AQC}, long=Adiabatic Quantum Computing}
\DeclareAcronym{adapt}{short=\textsc{ADAPT}, long=Adaptive Derivative-Assembled Pseudo-Trotter}
\DeclareAcronym{pva}{short=\textsc{PVA}, long=Projector Variationnal Ansatz}
\DeclareAcronym{pf}{short=\textsc{PF}, long=Product Formulas}
\DeclareAcronym{mpf}{short=\textsc{MPF}, long=Multi-Product Formulas}
\DeclareAcronym{ts}{short=\textsc{TS}, long=Taylor-Series}
\DeclareAcronym{neeqma}{short=\textsc{NEEQMA}, long=Numerical Error Extraction by Quantum Measurement Algorithm}
\DeclareAcronym{aqe}{short=\textsc{AQE}, long=Ancilla Quantum Encoding}
\DeclareAcronym{sqsp}{short=\textsc{SQSP}, long=Sparse Quantum State Preparation}
\DeclareAcronym{qlsp}{short=\textsc{QLSP}, long=Quantum Linear Systems Problem}

\DeclareAcronym{gas}{short=\textsc{GAS}, long=Grover's Adaptive Search}
\DeclareAcronym{ga}{short=\textsc{GA}, long=Grover's search Algorithm}
\DeclareAcronym{bop}{short=\textsc{BOP}, long=Binary Optimization Problem}
\DeclareAcronym{qspbo}{short=\textsc{QSPBO}, long=Quantum Signal Processing Based Oracle}
\DeclareAcronym{qabo}{short=\textsc{QABO}, long=Quantum Arithmetic Based Oracle}
\DeclareAcronym{qdbo}{short=\textsc{QDBO}, long=Quantum Dictionary Based Oracle}
\DeclareAcronym{ph}{short=\textsc{PH}, long=Problem Hamiltonian}
\DeclareAcronym{aa}{short=\textsc{AA}, long=Amplitude Amplification}
\DeclareAcronym{qse}{short=\textsc{QSE}, long=Quantum Signal Estimation} 
\DeclareAcronym{qso}{short=\textsc{QSO}, long=Quantum Signal Oraclization}
\DeclareAcronym{fpgs}{short=\textsc{FPGS}, long=Fixed-Point Grover Search}
\DeclareAcronym{qae}{short=\textsc{QAE}, long=Quantum Amplitude Estimation}
\DeclareAcronym{eaa}{short=\textsc{EAA}, long=Exact Amplitude Amplification}
\DeclareAcronym{qw}{short=\textsc{QW}, long=Quantum Walk}

\DeclareAcronym{nmr}{short=\textsc{NMR}, long=Nuclear Magnetic Resonance}
\DeclareAcronym{hpc}{short=\textsc{HPC}, long=High Performance Computing}
\DeclareAcronym{qpu}{short=\textsc{QPU}, long=Quantum Processing Unit}
\DeclareAcronym{gpu}{short=\textsc{GPU}, long=Graphics Processing Unit}
\DeclareAcronym{cpu}{short=\textsc{CPU}, long=Central Processing Unit}
\DeclareAcronym{msb}{short=\textsc{MSB}, long=Most Significant Bit}
\DeclareAcronym{fom}{short=\textsc{FOM}, long=Figures Of Merits}
\DeclareAcronym{kpi}{short=\textsc{KPI}, long=Key Performance Indicator}
\DeclareAcronym{nisq}{short=\textsc{NISQ}, long=Noisy Intermediate Scale Quantum computer}
\DeclareAcronym{isq}{short=\textsc{ISQ}, long=Intermediate Scale Quantum computer}
\DeclareAcronym{ftqc}{short=\textsc{FTQC}, long=Fault Tolerant Quantum Computer}
\DeclareAcronym{mbqc}{short=\textsc{MBQC}, long=Measurement Based Quantum Computing}

\DeclareAcronym{dft}{short=\textsc{DFT}, long=Discrete Fourier Transform}
\DeclareAcronym{hsp}{short=\textsc{HSP}, long=Hidden Sub-group Problem}
\DeclareAcronym{sat}{short=\textsc{SAT}, long=Boolean SATisfiability}
\DeclareAcronym{cnf}{short=\textsc{CNF}, long=Conjunctive Normal Form}
\DeclareAcronym{dnf}{short=\textsc{DNF}, long=Disjunctive Normal Form}
\DeclareAcronym{qubo}{short=\textsc{QUBO}, long=Quadratic Unconstrained Binary Optimization}
\DeclareAcronym{hubo}{short=\textsc{HUBO}, long=High order Unconstrained Binary Optimization}
\DeclareAcronym{hobo}{short=\textsc{HOBO}, long=High Order Binary Optimization}
\DeclareAcronym{cpbo}{short=\textsc{CPBO}, long=Constrained Polynomial Binary Optimization}
\DeclareAcronym{sto3g}{short=\textsc{STO3G}, long=Slater-Type Orbital represented by 3 gaussian}
\DeclareAcronym{ghz}{short=\textsc{GHZ}, long=Greenberger-Horne-Zeilinger}
\DeclareAcronym{ffft}{short=\textsc{FFFT}, long=Fast Fermionic Fourier Transform}
\DeclareAcronym{lep}{short=\textsc{LEP}, long=Low Energy Physics}
\DeclareAcronym{hep}{short=\textsc{HEP}, long=High Energy Physics}
\DeclareAcronym{pde}{short=\textsc{PDE}, long=Partial Differential Equations}
\DeclareAcronym{ode}{short=\textsc{ODE}, long=Ordinary Derivative Equations}
\DeclareAcronym{fdm}{short=\textsc{FDM}, long=Finite-Difference Method}
\DeclareAcronym{fem}{short=\textsc{FEM}, long=Finite-Element Method}
\DeclareAcronym{fvm}{short=\textsc{FVM}, long=Finite-Volume Method}
\DeclareAcronym{ic}{short=\textsc{IC}, long=Initial Conditions}
\DeclareAcronym{lc}{short=\textsc{LC}, long=Limit/Boundary Conditions}
\DeclareAcronym{adi}{short=\textsc{ADI}, long=Alternating Direction Implicit}
\DeclareAcronym{qml}{short=\textsc{QML}, long=Quantum Machine Learning}
\DeclareAcronym{ml}{short=\textsc{ML}, long=Machine Learning}
\DeclareAcronym{pca}{short=\textsc{PCA}, long=Principal Component Analysis}

\DeclareAcronym{hc}{short=\textsc{\hc}, long=Hermitian Conjugate}
\DeclareAcronym{svd}{short=\textsc{SVD}, long=Singular Value Decomposition}
\DeclareAcronym{svt}{short=\textsc{SVT}, long=Singular Value Transformation}
\DeclareAcronym{bch}{short=\textsc{BCH}, long=Baker–Campbell–Hausdorff}

\DeclareAcronym{qan}{short=\textsc{QA}, long=Quantum Annealing}

\renewcommand{\selectlanguage}[1]{}

\begin{document}


\title{\acl{hs} and \acl{lcu} Decomposition of Structured Matrices}

\def\orgadep{CEA, List, F-91120}
\def\orga{CEA}
\def\loc{Palaiseau, France} 
\def\univ{Université Paris-Saclay}

\author{Robin OLLIVE (0009-0006-7539-363X)}
\author{Stephane LOUISE (0000-0003-4604-6453)}
\affiliation{\textit{\univ}, \orgadep, \loc}


\date{\today}

\begin{abstract}
To process a problem with a \ac{qpu}, it must be transformed into a sequence of quantum operators, or gates. 
These operators are either packed into a query (\textit{i.e.} quantum algorithm primitive) that encodes the problem, or used to construct the cost function for \ac{vqa}.
Typical queries are the problem \ac{hs} and the problem \ac{bei}.
To construct the circuits associated  with the quantum description, the problem must be mapped as a \ac{lch} or a \ac{lcu} matrices.
All the summed Hamiltonian matrices or unitary matrices must have a known decomposition in basic gates.
The complexity of this query should be incorporated into the quantum algorithm's query complexity, thereby limiting the processing possibilities of \ac{qpu} for many problems.

Qubitization constructs a specific query that behaves like a single qubit when expressed in the appropriate basis.
In this work, we propose Hamiltonian matrices used to map the problem of interest, and that also behave like a single qubit when expressed in the appropriate basis.
It leads to a framework, the \ac{sth} Framework, able to implement most of the typical problems considered for quantum computing.
These methods address many problems implemented on \ac{qpu}s, ranging from second-quantization chemistry operators to graphs associated with \ac{pde}, sparse matrices, and higher-order optimization problems.
This work underlines interesting properties associated with the \ac{sth} basic gate decomposition.
These include the ability to switch between \ac{lch} and \ac{lcu}, map non-Hermitian problems, and construct the quantum circuit primitives (queries) required for quantum computing.
We also provide a list of \ac{sth} that are used for the matrix decomposition of many structured matrices. 
These structured matrices are associated with graph adjacency matrices that can be combined to implement structured matrices. 
\end{abstract}

\maketitle

\tableofcontents

\section{Introduction}
\paragraph{Context}
\acresetall
Quantum algorithms such as \ac{qpe} \cite{kitaev_quantum_1995}, \ac{hhl} \cite{harrow_quantum_2009}, and digital \ac{qan} \cite{farhi_quantum_2000} require the \ac{hs} of the problem.
Similarly, \ac{qsp} \cite{low_quantum_2017, low_hamiltonian_2019} and \ac{qsvt}-based algorithms \cite{gilyen_quantum_2019, martyn_grand_2021} require a signal operator constructed using \ac{bei}. 
Finally, \ac{vqa} \cite{peruzzo_variational_2014, yung_transistor_2014, wecker_progress_2015, mcclean_theory_2016} relies on measuring a cost function via the Hadamard test or via \ac{pvm}. 

The Hadamard test and the \ac{bei} are constructed by assembling the \ac{lcu} associated matrices: $ \qop{M_{p}} = \sum_{i} \beta_{i} \qop{U}_{i} $.
The \ac{hs} of the problem is often directly constructed from the exact \ac{hs} of the \ac{lch} summands: $ \qop{H_{p}} = \sum_{i} \alpha_{i} \qop{H_{i}} $, assembled by \ac{pf} \cite{ostmeyer_suzuki-trotter_2026, childs_theory_2021} \footnote{Note that in specific cases such as \ac{hsp} algorithm, \ac{hs} is constructed using Quantum Arithmetics \cite[section 7]{cleve_quantum_1998}, \cite{ossorio-castillo_generalisation_2023}, \cite[annex C]{ollive_quantum_2025}.}.
To implement matrix problems efficiently on quantum computers, it is necessary to know the basic gate decompositions of the problem that a quantum computer can execute.
Thus, to implement efficiently a matrix associated with a problem, it is required to know a decomposition into basic gates of as many standard Hamiltonian matrices as possible.
These matrices are then combined to construct the associated queries.

A major quantum information concept introduced by \cite{low_hamiltonian_2019} is qubitization.
This approach transforms an arbitrary \ac{bed} matrix into a qubitized operator that has a subspace in which it behaves similarly to a single qubit.
It provides an intuitive representation of the operation in the qubitized space, analogous to the same operation on a single qubit.
Particularly, it allows the extension of \ac{qsp} to many-qubit Hamiltonians.

\paragraph{Similar Work}
Two separate strategies were initially proposed to format matrices for quantum computing: the \ac{ps} matrix decomposition, generally obtained by direct mapping of chemistry or optimization problems, and the $d$-sparse matrices implemented by discreate time \ac{qw}.
While a few methods exist for implementing the \ac{hs} for other Hamiltonians, the recent increase in the number of Hamiltonians that can be implemented stems from the proper definition of the \ac{bei} concept, which was also introduced by \cite{low_hamiltonian_2019}.

Compatible with both \ac{hs} (by \ac{pf}) and \ac{bei} (as an \ac{lcu}), the \ac{ps} decomposition of the matrix of interest is a very attractive option as it theoretically allows the decomposition of any matrix.
This decomposition generally requires an exponential number of summands.
Many schemes have been proposed to optimize the prefactor of the exponential cost in \ac{ps} decomposition algorithms \cite{pesce_h2zixy_2021,koska_tree-approach_2024,jones_decomposing_2024,hantzko_tensorized_2024,georges_pauli_2025,philo_phase_2026}.
It is important to note that this decomposition is not required for problems that directly map to these matrices, such as chemistry \cite{whitfield_simulation_2011} or \ac{qubo} formulations.
To lower the cost of implementing \ac{ps}, several proposals aim to gather \ac{ps} terms that can be simultaneously diagonalized (\textit{a.k.a.} partitioning) to implement them at once \cite{berg_circuit_2020,mukhopadhyay_synthesizing_2023,martinez-martinez_assessment_2023}.
Diagonal matrices with a function that depends on the matrix component index, evaluated at all diagonal indices, are important in quantum computing, for instance, for state preparation.
Pauli decompositions that approximate such diagonal matrices are thus a natural subfield \cite{welch_efficient_2014,zylberman_efficient_2024,ayral_walsh_2026}.

After the \ac{bei} concept was established, a new sub-field emerged: research on problem-relevant unitary matrix gate decompositions.
These matrices are compatible only with \ac{lcu} but are generally not \ac{lch}, meaning they do not allow \ac{hs} via \ac{pf} or measurement via \ac{pvm} for the problem matrix.
Two illustrative cases are displacement matrices (circulant and related matrices) that do a permutation of the state indices.
Circulant matrices are primarily constructed using two strategies: one based on \ac{qa} \cite{wan_asymptotic_2018,wan_block-encoding-based_2021,camps_explicit_2023,sunderhauf_block-encoding_2024,sturm_efficient_2025,kharazi_explicit_2025,zecchi_block_2026}  and the other on the \ac{qft} as a basis change \cite{lubasch_quantum_2025,javanmard_boundary-aware_2026}, which diagonalizes them.
Another important example is the multiplication (and division) by two matrices used by \cite{camps_explicit_2023,sunderhauf_block-encoding_2024} to implement an extended binary tree.

The last family of operators that can be both \ac{hs} and \ac{be} is the \ac{scb} \cite{ollive_gate_2025} that, combine with the \ac{ps}, are naturally suited to implement practical problems.
The \ac{scb} was inspired by \cite{gilliam_grover_2021}, where an \ac{hhl} algorithm for \ac{hubo} Hamiltonians was constructed using $\qop{n}$ and $\qop{m}$ operators extended to fermionic ladder operator and sparse matrices.
Papers that derive circuits similar to \cite{ollive_gate_2025}'s \ac{scb} or that address similar problems mainly stem from efficient implementations of tridiagonal matrices.
The \ac{hs} of tridiagonal matrices (for processing \ac{pde}) by \cite{vazquez_enhancing_2022} leads to circuit similar that the one obtained following the \ac{scb} framework.
\cite{sato_hamiltonian_2024} notices that the different \ac{hs} terms (combined via \ac{pf}) in this specific Laplacian operator's decomposition can be factorized into a matrix tensor product of \ac{scb} operators.
The \ac{lcu} decomposition of tridiagonal matrices has led to work by \cite{liu_variational_2021,gnanasekaran_efficient_2024} that implements \ac{scb} operators (called the sigma basis) by using \ac{lcu} to measure (by Hadamard test) multidiagonal matrices in the \ac{vqls} algorithm.
In \cite{gnanasekaran_efficient_2026}, this approach is then extended to \ac{be} of tensor products involving \ac{scb} and other unitaries (referred to as linear combinations of 'things').
They also provide a decomposition algorithm for arbitrary matrices using \ac{scb}.
A circuit construction based on a similar idea is proposed in \cite{simon_ladder_2025} for encoding fermionic ladder operators.
It is interesting to note that the unitary matrices proposed in \cite{liu_variational_2021,gnanasekaran_efficient_2024,gnanasekaran_efficient_2026} have a design based on \ac{lcu} that is not compatible with \ac{hs} via \ac{pf} or expectation value evaluation via \ac{pvm}.


The most efficient \ac{lch} (both a \ac{lcu} and an \ac{lch}) decomposition of tridiagonal circulant matrices was proposed by \cite{vazquez_enhancing_2022,ty_double-logarithmic_2024}. Although the gate structure presented in their work is efficient, it is not described within the \ac{scb} formalism.
The \ac{sth} framework introduced in this paper was developed to unify all the aforementioned methods as specific instances of a single, comprehensive framework. This framework enables the implementation of matrices as both \ac{lch} and \ac{lcu} summand decompositions, compatible with quantum computing requirements.

\paragraph{Contributions}
This article extends the \ac{ps} and \ac{scb} Hamiltonians to the \ac{sth}, a family of operators well-suited for encoding problems on quantum computers.
These Hamiltonians have a subspace in which they behave following single-qubit physics similarly than unitary (\textit{i.e.} signal operator) constructed by qubitization.
The first section, \Cref{section_def} explains how to construct the previously mentioned primitives with the \ac{sth}.
The second section, \Cref{sec_fundamental_examples} provides a new, simpler, and unified description of \ac{ps} and \ac{scb} Hamiltonians within this framework.
Finally, the third section, \Cref{sec_well_structured_problem} provides a list of \ac{sth} relevant for mapping structured matrices associated with a tractable, well-characterized decomposition using basic quantum gates.
The matrix compositions in this list represent the other quantum information matrices proposed in the previously presented state of the art.
The graphs associated with these structured adjacency matrices are also provided.

\section{Special Tripotent Hamiltonian Operators Definition \label{section_def}}
\ac{sth}\footnote{In the preliminary preprint of this paper, the name Qubitized Hamiltonian was used for this framework.
It appears that this name is confusing, as for authors such as \cite{pocrnic_improved_2026}, qubitized Hamiltonians refer to any \ac{bed} Hamiltonian that is transformed into a signal operator by qubitization.} form a sub-family of Hamiltonian matrices that are directly expressed as single qubits in their eigenbasis:
\begin{equation}
\begin{aligned}
  \qop{H_{i}} & = \sum _{k = 0}^{\taille{\qop{H_{i}}} - 1} \lambda_{k} \ketbra{\lambda_{k}} 
  = \qop{\Pi}_{\ket{\lambda}}  - \qop{\Pi}_{\ket{\lambda_{\perp}}} \\
  & = \qop{V_{i}}^{\dag} \cdot \qop{\lambda_{i}} \cdot \qop{V_{i}} 
  = \qop{n_{\ket{\perp}} Z}\{ \ket{\lambda}; \ket{\lambda_{\perp}} \} \\
  & = \begin{bmatrix}
      0 & 0 & 0 \\
      0 & 1 & 0 \\
      0 & 0 & - 1
  \end{bmatrix}
  \begin{matrix}
      \aide{\ket{\perp}} \\
      \aide{\ket{\lambda}} \\
      \aide{\ket{\lambda_{\perp}}}
  \end{matrix} 
  \label{eq_def_qubitized_hamiltonian}
\end{aligned}
\end{equation}
with $ \ket{\lambda}, \ket{\lambda_{\perp}}, \ket{\perp} $ an arbitrary eigenstates $\ket{\lambda_{k}}$ with a degenerated eigenvalue respectively equal to one, minus one, and zero.
The non-zero eigenspace is decomposed by the commuting orthogonal projectors:
\begin{equation}
    \qop{\Pi}_{\ket{\lambda}} = \sum_{k |\lambda_{K} = 1} \ketbra{\lambda_{k}} \text{, and: } \qop{\Pi}_{\ket{\lambda_{\perp}}} = \sum_{k |\lambda_{K} = -1} \ketbra{\lambda_{k}}
\end{equation}
A third projector allows to describes the complete quantum system space with the state whose eigenvalues are zero:
\begin{equation}
    \qop{\Pi}_{\ket{\perp}} = \sum_{k |\lambda_{K} = 0} \ketbra{\lambda_{k}}
\end{equation}

$\qop{\lambda_{i}}$ is a diagonal matrix and $ \qop{V_{i}} = \sum_{k} \ket{\binaire{k}} \bra{\lambda_{k}} $ is a unitary matrix that transform an eigenbasis into the computational basis.
The notation $\qop{n_{\ket{\perp}} Z}\{ \ket{\lambda}; \ket{\lambda_{\perp}} \}$ is introduced in \cite{ollive_gate_2025} to indicate the subspace in which the tripotent Hamiltonian behave like a $\qop{Z}$ gate\footnote{The notation $ \qop{n_{\ket{\perp}} U}\{ \ket{\lambda}; \ket{\lambda_{\perp}} \} = \qop{U}\{ \ket{\lambda}; \ket{\lambda_{\perp}} \} $ means that the eigenvalues associated to $\qop{\Pi}_{\ket{\perp}}$'s states equal to zero.
It is similar to gate controlled by a set of states: $ \qop{C_{\ket{\perp}} U}\{ \ket{\lambda}; \ket{\lambda_{\perp}} \} = \qop{\Pi}_{\ket{\perp}} + \qop{U}\{ \ket{\lambda}; \ket{\lambda_{\perp}} \} $ whose state eigenvalues equal to one.}.
More formally, the tripotent Hamiltonian family contains all the Hermitian matrices with the eigenvalue included in $ \lambda_{k} \in \{ -1, 0, 1 \} $ which imply this matrix are tripotent: $ \qop{H}_{i}^{3} = \qop{H}_{i} $.
The \ac{sth} have an extra condition:
it can be expressed using a single $n$-controlled $Z$-rotation gate: \begin{equation}
    \qop{H}_{i} = \qop{V_{i}}^{\dag} \cdot \qop{n_{\ket{\perp}} Z}\{ \ket{0}_{j}, \ket{1}_{j}\} \cdot \qop{V_{i}}
\end{equation}
where $j$ is the index of one of the qubits whose basis is $\{ \ket{0}, \ket{1} \}$.
The \ac{sth} thus have a trace equals to zero, a non-zero eigenspace size equals to a power of two, and the central operator can be factorised as: 
\begin{equation}
    \qop{n_{\ket{\perp}} Z}\{ \ket{0}_{j}, \ket{1}_{j}\} = \bigotimes_{j = 0}^{nb_{qb} - 1} \qop{PSCB}_{j}
\end{equation}
with $ \qop{PSCB}_{j} \in \{ \qop{n}, \qop{m}, \qop{I}, \qop{Z} \} $ diagonal matrices:
\begin{equation}
    \qop{Z} = 
    \begin{bmatrix}
        1  & 0 \\
        0 & -1
    \end{bmatrix} ;
    \qop{n} = 
    \begin{bmatrix} ;
        0 & 0 \\
        0 & 1
    \end{bmatrix} ;
    \qop{m} = 
    \begin{bmatrix}
        1 & 0 \\
        0 & 0
    \end{bmatrix}
\end{equation}
It is worth noting that other tripotent Hamiltonians implemented using multiple rotations (\textit{i.e.} a sum of \ac{sth}) exist but are not implemented as a single summand (with a single rotation) thanks to the proposed formalism.
An illustrative example is: $ \qop{V_{i}}^{\dag} \cdot ( \qop{m} \qop{m} \qop{Z} + \qop{n} \qop{Z} \qop{Z} ) \cdot \qop{V_{i}} $.
What we are searching for is a small number of \ac{sth} with a known, easy-to-implement basis change $\qop{V}_{i}$ to describe useful matrices.

The most popular example included in the \ac{sth} family is the tensor product of Pauli matrices\footnote{
    Pauli matrices are both Hermitian and unitary matrices (more specifically reflection matrices).
} (plus identity):
\begin{equation}
    \qop{X} = 
    \begin{bmatrix}
        0 & 1 \\
        1 & 0
    \end{bmatrix} ;
    \qop{Y} = 
    \begin{bmatrix}
        0  & -\ii \\
        \ii & 0
    \end{bmatrix} ;
    \qop{Z} = 
    \begin{bmatrix}
        1  & 0 \\
        0 & -1
    \end{bmatrix}
\end{equation}
the \ac{ps}:
\begin{equation}
    \qop{PS} = \bigotimes_{j = 0}^{nb_{qb} - 1} \qop{P}_{j}
\end{equation}
with $ \qop{P}_{j} \in \{ \qop{I}, \qop{X}, \qop{Y}, \qop{Z} \} $.
The $4^{N}$ \ac{ps} are a sufficient basis to map all the matrices of size $2^{N}$ ($N$ qubits) \cite[eq 4.114]{nielsen_quantum_2010}, \cite[II.C]{koska_tree-approach_2024}.
It implies that \ac{sth} form a redundant family of generating matrices.
It is nevertheless interesting to be able to construct as many usual matrices as possible thus allowing a better expressivity to map the matrix associated with the problem of interest.
An illustrative example is the implementation of a single component of a matrix (plus its Hermitian conjugate ($\mathit{h.c.}$) so that the matrix is Hermitian).
It needs $2^{N}$ \ac{ps} but only one tensor product of \ac{scb} matrix \cite[section V.D]{ollive_gate_2025}:
\begin{equation}
    \qop{n} = 
    \begin{bmatrix}
        0 & 0 \\
        0 & 1
    \end{bmatrix} ;
    \qop{m} = 
    \begin{bmatrix}
        1 & 0 \\
        0 & 0
    \end{bmatrix} ;
    \qop{\sigma} = 
    \begin{bmatrix}
        0 & 1 \\
        0 & 0
    \end{bmatrix} 
\end{equation}
a \ac{scb}-string:
\begin{equation}
    \qop{SCB} = \bigotimes_{j = 0}^{nb_{qb} - 1} \qop{A}_{j} + \qop{\hc}
\end{equation}
with $ \qop{A}_{j} \in \{ \qop{n}, \qop{m}, \qop{\sigma}, \qop{\sigma}^{\dag} \} $ and at least one value of $j$ such that $ \qop{A}_{j} \in \{ \qop{\sigma}, \qop{\sigma}^{\dag} \} $.
One can recognise the number of excitation ($\qop{n}$ and $\qop{m}$) and the ladder operators ($\qop{\sigma}$ and $\qop{\sigma}^{\dag}$) from second quantization.
It has already been proven that the mix of these two basis leads to a family of generating matrices that are well-suited at mapping second-quantization Hamiltonians, higher-order optimization problems and sparse matrices in \cite{ollive_gate_2025}.
It is shown in \Cref{sec_fundamental_examples} how these matrices are constructed more straightforwardly from the \ac{sth} framework presented here.
Another section, \Cref{sec_well_structured_problem} extends this family with structured graphs that can be implemented thanks to \ac{sth}.

It is important to note that the Hamiltonian such as:
\begin{equation}
    \qop{H_{i}} = \bigotimes_{j = 0}^{nb_{qb} - 1} \qop{PSCB}_{j} =
    \begin{bmatrix}
      0 & 0 \\
      0 & 1
  \end{bmatrix}
  \begin{matrix}
      \aide{\ket{\perp}} \\
      \aide{\ket{\lambda}} \\
  \end{matrix}
  = \qop{n}\{ \ket{\perp}; \ket{\lambda} \}
  = \qop{\Pi}_{\ket{\lambda}}
\end{equation}
or more generally:
\begin{equation}
    \qop{\Pi}_{\qop{V}_{i} \ket{\lambda}} = \qop{V}_{i}^{\dag} \cdot \qop{\Pi}_{\ket{\lambda}} \cdot \qop{V}_{i}
\end{equation}
with $ \qop{PSCB}_{j} \in \{ \qop{n}, \qop{m}, \qop{I} \} $ are tripotent Hamiltonians but also are included in a subfamily referred to as idempotent Hamiltonian (that include projectors) which respect: $ \qop{H}_{i}^{2} = \qop{H}_{i} $.
These projectors follow algebraic properties that can be derived similarly to the \ac{sth} one but more straightforwardly.
These properties are thus also detailed when they differ from \ac{sth}.

\subsection{\acl{sth} Reduction to One Qubit}

\paragraph{\acl{sth} and Bloch-Sphere}
As mentioned in \Cref{section_def}, the trace of the \ac{sth} equals zero: $ \mathrm{Tr}[\qop{H_{i}}] = 0 $ and the eigenstates are gathered in two degenerate groups with non-zero eigenvalues.
Thus $ \mathrm{Det}[e^{\ii t \qop{H_{i}}}] = 1 $ is a matrix in the special unitary group.
It implies that \ac{sth} \ac{hs} applied on a state $\ket{\psi}$ can be represented by a Bloch-sphere like representation whose $z$ axis othogonal vectors are $ \{ \qop{\Pi}_{\ket{\lambda}} \ket{\psi}, \qop{\Pi}_{\ket{\lambda_{\perp}}} \ket{\psi} \} $.
This links qubit representations and \ac{sth}.

\begin{figure}[tb]
    \begin{center}
    \resizebox{\linewidth}{!}
    {\includegraphics{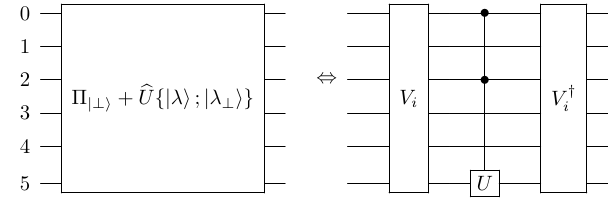}}
    \end{center}
    \caption{Application of the one-qubit gate $\qop{U}$ in the non-zero subspace $ \{ \ket{\lambda}, \ket{\lambda_{\perp}} \} $.
    Here, $ \qop{V_{i}} = \ket{0}^{\otimes 2}_{0, 2} \left(  \ket{0}_{5} ( \sum_{k | \lambda_{k} = 1} \ket{\binaire{k}}_{1, 3, 4} \bra{\lambda_{k}}) \right. + \left. \ket{1}_{5} ( \sum_{k | \lambda_{k} = -1} \ket{\binaire{k}}_{1, 3, 4} \bra{\lambda_{k}}) \right) + \sum_{k | \ \lambda_{k} = 0} \ket{\binaire{k}} \bra{\lambda_{k}} $.
    In this example, the non-zero eigenspace is of size $2^{4}$.}
    \label{fig_qc_basic}
\end{figure}

\paragraph{Reminder Projector-Controlled-One-Qubit Gate}
Projector-controlled phase shifts are objects introduced by \cite[eq 28]{martyn_grand_2021} that permit the application of a phase gate on a qubitized \ac{bei} (signal operator):
\begin{equation}
 \qop{\Pi}_{\phi} = e^{i \phi \qop{Z}} \otimes \sum_{k} \ketbra{\lambda_{k}}
\end{equation}
By analogy, we propose the Projector-controlled-one-qubit-gate that permits the application of a phase (or a Pauli Z) gate on a \ac{sth} non-zero eigenspace.
If the projector-controlled-one-qubit-gate decomposition is known, the gate can efficiently be constructed, as it is a specific basis transformation that projects the degenerated non-zero eigenspace on a specific qubit:
\begin{equation}
\begin{aligned}
    \qop{V_{i}}^{\dag} & = \left( (\sum_{k |\lambda_{K} = 1}  \ket{\lambda_{k}} \bra{\binaire{k}}_{p}) \bra{0}_{j} \right. \\
    & \qquad \left. + (\sum_{k |\lambda_{K} = -1} \ket{\lambda_{k}} \bra{\binaire{k}}_{p}) \bra{1}_{j} \right) \bra{0}^{\otimes n_{i}}_{r} \\
    & \quad + \sum_{k | \lambda_{k} = 0} \ket{\lambda_{k}} \bra{\binaire{k}}_{rjp}
\end{aligned}
\end{equation}
with $j$ the qubit that controls the rotation of the non-zero eigenspace, $r$ the qubits that control this rotation, $p$ the other qubits of the system, and $\ket{\binaire{k}}$ a state of the computational basis associated to the vector subspace.
The $\qop{V_{i}}$ matrix is not uniquelly defined if the non-zero subspace contains more than two states: the degenerated subspace can be decomposed with different basis and transformed (by $\qop{V}_{i}$) on different state of the computational basis.
The qubit associated with the different eigenvalue projection is referred to as the reduct-qubit.
It implies that a one-qubit gate $\qop{U}$ can be applied on the non-zero subspace with the quantum circuit \Cref{fig_qc_basic}: 
\begin{equation}
\begin{aligned}
    \qop{C_{\ket{\perp}} U}\{ \ket{\lambda}; \ket{\lambda_{\perp}} \} & = \qop{V_{i}}^{\dag} \cdot \qop{C_{\ket{0}_{r}^{\otimes n_{i}}} U}\{ \ket{0}_{j}; \ket{1}_{j} \} \cdot \qop{V_{i}} \\
    & = \qop{\Pi}_{\ket{\perp}} + \qop{U}\{ \ket{\lambda}; \ket{\lambda_{\perp}} \} \\
    & = \begin{bmatrix}
        1 & 0 & 0 \\
        0 & U_{1, 1} & U_{1, 2} \\
        0 & U_{2, 1} & U_{2, 2}
    \end{bmatrix}
    \begin{matrix}
        \aide{\ket{\perp}} \\
        \aide{\ket{\lambda}} \\
        \aide{\ket{\lambda_{\perp}}}
    \end{matrix} 
\label{eq_qubitized_gate}
\end{aligned}
\end{equation}
Note that the off-diagonal terms ($U_{1, 2}$ and $U_{2, 1}$) are well-defined only when the non-zero eigenspace has size two and else depend on the $\qop{V}_{i}$ basis change matrix.

\subsubsection{Tripotent Hamiltonian Tensor Product \label{subsec_qubitized_tensor_product}}
\begin{figure}[tb]
    \begin{center}
        \resizebox{\linewidth}{!}
        {\includegraphics{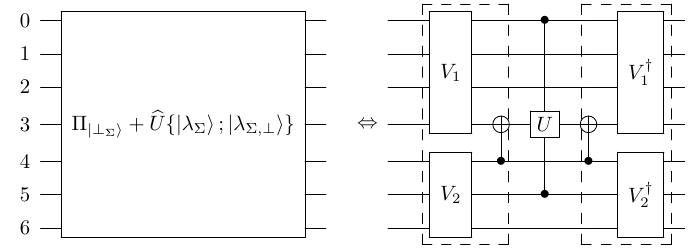}}
    \end{center}
    \caption{Application of the one qubit gate $\qop{U}$ in the non-zero eigenspace $ \{ \ket{\lambda_{\Sigma}}, \ket{\lambda_{\perp, \Sigma}} \} $.
    The new projector-controlled-one-qubit-gate $\qop{V_{\Sigma}}^{(\dag)}$ (and its conjugate) are the gates in the dashed box.
    Here, $ \qop{V_{1}} = \ket{0}_{0} \left( \ket{0}_{3} ( \sum_{k | \lambda_{k} = 1} \ket{\binaire{k}}_{1, 2} \bra{\lambda_{k}} ) \right. + \left. \ket{1}_{3} ( \sum_{k | \lambda_{k} = -1} \ket{\binaire{k}}_{1, 2} \bra{\lambda_{k}} ) \right) + \sum_{k | \lambda_{k} = 0} \ket{\binaire{k}} \bra{\lambda_{k}} $
    and $ \qop{V_{2}} = \ket{0}_{5} \left( \ket{0}_{4} ( \sum_{k | \lambda_{k} = 1} \ket{\binaire{k}}_{6} \bra{\lambda_{k}}) \right. + \left. \ket{1}_{4} ( \sum_{k | \lambda_{k} = -1} \ket{\binaire{k}}_{6} \bra{\lambda_{k}}) \right) + \sum_{k | \lambda_{k} = 0} \ket{\binaire{k}} \bra{\lambda_{k}} $.
    }
    \label{fig_qc_tensor_product}
\end{figure}

Combining the projector-controlled-one-qubit-gates of two (or more) \ac{sth} via a tensor product in a single \ac{sth} can be done efficiently.
The tensor product of two tripotent Hamiltonians stays a tripotent Hamiltonian:
\begin{equation}
\begin{aligned}
    \qop{H}_{\Sigma} & = \qop{H}_{1} \otimes \qop{H}_{2} \\
    & = (\qop{\Pi}_{\ket{\lambda_{1}}} - \qop{\Pi}_{\ket{\lambda_{\perp, 1}}}) \otimes (\qop{\Pi}_{\ket{\lambda_{2}}} - \qop{\Pi}_{\ket{\lambda_{\perp, 2}}}) \\
    & = \qop{\Pi}_{\ket{\lambda_{\Sigma}}} - \qop{\Pi}_{\ket{\lambda_{\perp, \Sigma}}}
\end{aligned}
\end{equation}
with $ \qop{\Pi}_{\ket{\lambda_{\Sigma}}} = \qop{\Pi}_{\ket{\lambda_{1}}} \qop{\Pi}_{\ket{\lambda_{2}}} + \qop{\Pi}_{\ket{\lambda_{\perp, 1}}} \qop{\Pi}_{\ket{\lambda_{\perp, 2}}} $ and $ \qop{\Pi}_{\ket{\lambda_{\perp, \Sigma}}} = \qop{\Pi}_{\ket{\lambda_{1}}} \qop{\Pi}_{\ket{\lambda_{\perp, 2}}} + \qop{\Pi}_{\ket{\lambda_{\perp, 1}}} \qop{\Pi}_{\ket{\lambda_{2}}} $.
It is only required to combine the eigenspace by reporting the parity of the reduct-qubit on a chosen reduct-qubit.
It is done in \Cref{fig_qc_tensor_product} using a ladder (or string) of control-X gate that reverses the chosen reduct-qubit (here the $3$\textsuperscript{th} qubit) depending on the parity of the other reduct-qubit (here the $4$\textsuperscript{th} qubit):
\begin{equation}
\begin{aligned}
    \qop{V_{\Sigma}}^{\dag} & = ( \qop{V_{1}}^{\dag} \otimes \qop{V_{2}}^{\dag} ) \cdot \qop{C_{4} X_{3}} \\
\end{aligned}
\end{equation}

\paragraph{Tensor Product with Projectors}
This property extends to the combination of non-zero eigenspace with a projector:
\begin{equation}
\begin{aligned}
    \qop{H}_{\Sigma'} & = \qop{H}_{1} \otimes \qop{\Pi}_{\qop{V}_{2} \ket{\lambda_{2}}} \\
    & = (\qop{\Pi}_{\ket{\lambda_{1}}} - \qop{\Pi}_{\ket{\lambda_{\perp, 1}}}) \otimes \qop{\Pi}_{\qop{V}_{2} \ket{\lambda_{2}}} \\
    & = \qop{\Pi}_{\ket{\lambda_{1}}} \qop{\Pi}_{\qop{V}_{2} \ket{\lambda_{2}}} - \qop{\Pi}_{\ket{\lambda_{\perp, 1}}} \qop{\Pi}_{\qop{V}_{2} \ket{\lambda_{2}}}
\end{aligned}
\end{equation}
We can see that the non-zero eigenspace is recognized thanks to a control of the projector eigenvalue, it is thus not required to report the parity on the reduct-qubit of $\qop{H}_{1}$:
\begin{equation}
    \qop{V_{\Sigma'}}^{\dag} = \qop{V_{1}}^{\dag} \otimes \qop{V_{2}}^{\dag}
\end{equation}

\subsection{One- and Two- Qubit Like Algebraic Properties}
The single-qubit algebraic properties has the advantage of processing small matrices ($ 2 \times 2 $) with an associated three-dimensional representation which simplifies the computation of gate effects.
The following section uses this representation to illustrate some properties of \ac{sth}.

\subsubsection{\acl{sth} \acl{hs}}
The \ac{hs} of \ac{sth} is expressed as \cite[Section III]{ollive_gate_2025}:
\begin{equation}
\begin{aligned}
    e^{\ii t \qop{H_{i}}} & = \qop{\Pi}_{\ket{\perp}} + e^{\ii t} \qop{\Pi}_{\ket{\lambda}} + e^{-\ii t} \qop{\Pi}_{\ket{\lambda_{\perp}}} \\
    & = \qop{\Pi}_{\ket{\perp}} + \qop{R_{Z}}\{ \ket{\lambda}; \ket{\lambda_{\perp}} \}(2 t)
\end{aligned}
\end{equation}
It is a specific case of \Cref{eq_qubitized_gate} where $\qop{U} = \qop{R_{Z}}(2 t) $.
\paragraph{Projector \acl{hs}}
The Hamiltonian simulation of projectors is efficiently constructed using $ \qop{U} = \qop{P}(t) $.

\subsubsection{From \acl{lch} to \acl{lcu} \label{subsec_lch_to_lcu}}
\begin{figure}[tb]
    \begin{center}
        {\includegraphics{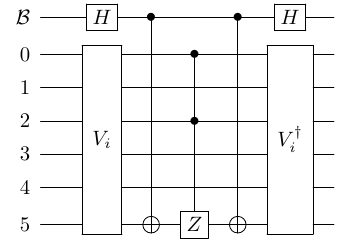}}
    \end{center}
    \caption{\ac{bei} of \Cref{fig_qc_basic}'s $\qop{H_{i}}$ when the first qubit reads $\ket{0}_{\mathcal{B}}$.}
    \label{fig_qc_be}
\end{figure}

In general, tripotent Hamiltonians are not unitary matrices, which means $ \ket{\perp} \neq \emptyset $.
To \ac{be} a tripotent Hamiltonian, one needs to find a sum of unitary matrices that subtract $\qop{\Pi}_{\ket{\perp}}$ to $ \qop{C_{\ket{\perp}} Z}\{ \ket{\lambda}; \ket{\lambda_{\perp}} \} $.
\cite[Section IV]{ollive_gate_2025} proposes a strategy to do it using a maximum of six unitary matrices by the tensor product of \ac{scb} and \ac{ps}.
This strategy is not optimal, it is possible to use only two unitary matrices to decompose a \ac{sth} using:
\begin{equation}
\begin{aligned}
    \qop{H}_{i} = \frac{\qop{C_{\ket{\perp}} Z}\{ \ket{\lambda}; \ket{\lambda_{\perp}} \} - \qop{C_{\ket{\perp}} (-Z)}\{ \ket{\lambda}; \ket{\lambda_{\perp}} \}}{2}
\end{aligned}
\label{eq_lcu_from_lch}
\end{equation},
with $ - \qop{Z} = \qop{X} \cdot \qop{Z} \cdot \qop{X} $.
\Cref{eq_lcu_from_lch}, is the two qubit equivalent of $ \qop{n} \qop{Z} = \frac{\qop{C Z} - \qop{C (-Z)}}{2} $.
A quantum circuit that constructs a \ac{bei} thanks to this technique is illustrated in \Cref{fig_qc_be}.

\paragraph{Projector \acl{be}}
It is possible to follow a similar strategy for the projectors:
\begin{equation}
    \qop{n}\{ \ket{\perp}; \ket{\lambda} \} = \frac{\qop{I} - \qop{Z}\{ \ket{\perp}; \ket{\lambda} \}}{2}
    \label{eq_projecteur_be}
\end{equation}

\textbf{To summarize this important subsection:} There exists a decomposition for Tripotent Hermitian into two involutions (and so unitary) matrices.
Idempotent Hamiltonian (orthogonal projectors) are a subfamily of the tripotent Hermitian and can thus be decomposed the same way.
In this paper, examples where these unitary gates are known quantum gates are provided.

\subsubsection{Naturally suited to Construct Signal Operator by Qubitization}
Qubitization of a \ac{bei} is described with two main steps \cite{low_hamiltonian_2019}:
\begin{itemize}
    \item The \ac{bei} is transformed into an involution that is still a \ac{bei} of the same problem:
    \begin{equation}
    \begin{aligned}
        \qop{I} & = \qop{S_{H_{p}}}^{2} \\
        \qop{H_{p}} & = \bra{\binaire{0}}_{\mathcal{B}} \bra{+}_{\mathcal{B}_{2}} \qop{S_{H_{p}}} \ket{+}_{\mathcal{B}_{2}} \ket{\binaire{0}}_{\mathcal{B}}
    \end{aligned}
    \end{equation}
    \item This involution is followed by another involution that constructs the signal operator:
    \begin{equation}
        \qop{W_{Y, H_{p}}} = \qop{S_{\psi_{i}}} \cdot \qop{S_{H_{p}}}
    \end{equation}
\end{itemize}
This subsection aims to show that the unitary matrices used to decompose the \ac{sth} directly lead to a \ac{bei} that is actually an involution.
This involution is directly constructed using the Prep-Sel-Prep routine:
\begin{equation}
\begin{aligned}
    \qop{S_{H_{p}}} & = \qop{PREP}^{\dag} \cdot \qop{Sel} \cdot \qop{PREP} \\
    \qop{Sel} & = \sum_{i} \ketbra{\binaire{i}} \otimes \qop{U_{i}} \\
    \qop{PREP} & = ( \sum_{i} \sqrt{\frac{|\alpha_{i}|}{\gamma ||\qop{H_{p}}||}} \ket{\binaire{i}} ) \bra{\binaire{0}} + (\sum_{i > 0} \ket{\varphi_{i}} \bra{\binaire{i}} )
\end{aligned}
\end{equation}
with $ \ket{\varphi_{i}} $ a state vector, $ \ket{\binaire{i}} $ the $i^{\text{th}}$ state of the computational basis, and $ \gamma ||\qop{H_{p}}|| $ an upper bound of the 1-norme (for instance: $ \gamma ||\qop{H_{p}}|| = \sum_{i} |\alpha_{i}| $).
When the summands are involutions, the construction of $\qop{S_{H_{p}}}$ is direct \cite[see Corollary 9]{low_hamiltonian_2019}:
\begin{equation}
\begin{aligned}
    \qop{S_{H_{p}}}^{2} & = \sum_{i} \ketbra{\binaire{i}} \otimes \qop{U_{i}}^{2} \otimes \qop{I} \\
    & = \qop{I} \text{ if } \qop{U_{i}}^{2} = \qop{I}
\end{aligned}
\end{equation}
This property arises from the unitary matrices used in \Cref{eq_lcu_from_lch} are square-roots of identity: $ ( \qop{C_{\ket{\perp}} Z}\{ \ket{\lambda}; \ket{\lambda_{\perp}} \} )^{2} = \qop{I} $.
It is naturally true for the unitary that derives from \Cref{eq_lcu_from_lch} and saves a \ac{be} repetition to construct the signal operator following \cite[the Proof of Lemma 10]{low_hamiltonian_2019}. 

\subsubsection{\acl{hc} Suppression \label{section_non_hermitian}} 
Some quantum algorithms process non-Hermitian matrices.
This section explains how to decompose, with unitary matrices, the state transition between the non-zero eigenstates of a \ac{sth}:
\begin{equation}
\begin{aligned}
    \qop{n_{\ket{\perp}} \sigma}\{ \ket{\lambda}; \ket{\lambda_{\perp}} \} & = \frac{\qop{C_{\ket{\perp}} X}\{ \ket{\lambda}; \ket{\lambda_{\perp}} \} - \ii \qop{C_{\ket{\perp}} Y}\{ \ket{\lambda}; \ket{\lambda_{\perp}} \}}{2} \\
    \qop{n_{\ket{\perp}} \sigma^{\dag}}\{ \ket{\lambda}; \ket{\lambda_{\perp}} \} & = \frac{\qop{C_{\ket{\perp}} X}\{ \ket{\lambda}; \ket{\lambda_{\perp}} \} + \ii \qop{C_{\ket{\perp}} Y}\{ \ket{\lambda}; \ket{\lambda_{\perp}} \}}{2} \\
\end{aligned}
\end{equation}
These equations are the two qubit equivalent of $ \qop{n} \qop{\sigma} = \frac{\qop{CX} + \ii \qop{C Y}}{2} $ and $ \qop{n} \qop{\sigma}^{\dag} = \frac{\qop{CX} - \ii \qop{C Y}}{2} $ with $ \pm \ii \qop{Y} = \qop{R_{Y}}(\mp \pi) $.
These techniques allow us to construct non-symmetric matrices. 
This means that it is possible to implement an arbitrary single component without its associated Hermitian conjugate which is critical to implement non-symmetric circulant matrices. 

\subsection{Measurement Circuit}
\begin{figure}[tb]
    \begin{center}
        {\includegraphics{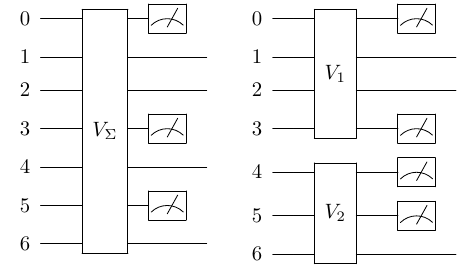}}
    \end{center}
    \caption{The two proposed measurement circuits allow to evaluate the expectation value: $ \bra{\psi} \qop{H_{\Sigma}} \ket{\psi} $ from \Cref{fig_qc_tensor_product}.
    Using the left circuit: $ \bra{\psi} \qop{H_{\Sigma}} \ket{\psi} = |\bra{\psi} \ket{0}_{3} \otimes \ket{11}_{0, 5}|^{2} - |\bra{\psi} \ket{1}_{3} \otimes \ket{11}_{0, 5}|^{2} $. \\
    Using the right circuit: $ \bra{\psi} \qop{H_{\Sigma}} \ket{\psi} = |\bra{\psi} \ket{00}_{3, 4} \otimes \ket{11}_{0, 5}|^{2} + |\bra{\psi} \ket{11}_{3, 4} \otimes \ket{11}_{0, 5}|^{2} - |\bra{\psi} \ket{01}_{3, 4} \otimes \ket{11}_{0, 5}|^{2} - |\bra{\psi} \ket{10}_{3, 4} \otimes \ket{11}_{0, 5}|^{2} $. \\
    This second circuit needs slightly more complex classical post-processing (\textit{i.e.} checking the hamming weight parity).
    However, only a shallower change of basis is required. 
    }
    \label{fig_qc_measurement}
\end{figure}

Many quantum algorithms, especially in the \ac{nisq} regime, require evaluating the expectation value of the Hamiltonian that encodes the problem by \ac{pvm} on a specific state $\ket{\psi}$.
To do so, an estimator is constructed with the sum of the summands' expectation value:
\begin{equation}
    \bra{\psi} \qop{H_{p}} \ket{\psi} = \sum_{i} \alpha_{i} \bra{\psi} \qop{H_{i}} \ket{\psi}
\end{equation}
Each summands' expectation values can be sampled statistically:
\begin{equation}
\begin{aligned}
    \bra{\psi} \qop{H_{i}} \ket{\psi} & = \sum_{k | \lambda_{k} = 1} |\braket{\lambda_{k} | \psi}|^{2} - \sum_{k | \lambda_{k} = -1} |\braket{\lambda_{k} | \psi}|^{2} \\
    & = \lim_{N \to \infty} \frac{n_{+} - n_{-}}{N}
\end{aligned}
\end{equation}
with $N$ the total number of shots and $n_{\pm}$ the number of shots associated to a $ \lambda_{k} = \pm 1 $ measurement.
An example of a measurement circuit to evaluate a \ac{sth} is described in \Cref{fig_qc_measurement}.
This measurement circuit does not require the central multicontrolled rotation gate needed to construct the other queries.
As a consequence, it can be used in \ac{nisq} algorithms when the state transition used for the change of basis is shallow.

\section{Fundamental example of Special Tripotent Hamiltonian Quantum Circuits \label{sec_fundamental_examples}}
This section quickly derives the well known \ac{ps} as well as the \ac{scb} \ac{hs} using the formalism presented in the previous section.
The objectives are both to propose warm-up examples of the \ac{sth} formalism and to remind the gate decomposition of operators (results of \cite{ollive_gate_2025}) that are used in \Cref{sec_well_structured_problem} to implement new matrices.

As explained in \Cref{subsec_lch_to_lcu}, a basic \ac{lcu} of \ac{sth} \ac{hs} allows to express the \ac{sth}.

\subsection{\acl{ps}}
\begin{figure}[tb]
    \begin{center}
        \resizebox{\linewidth}{!}
        {\includegraphics{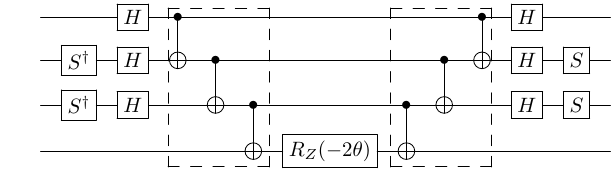}}
    \end{center}
    \caption{
    Decomposition of the unitary gate: $ e^{- \ii \theta \qop{X} \qop{Y} \qop{Y} \qop{Z}} $.
    }
    \label{fig_qc_hs_ps}
\end{figure}

The basis change that diagonalize the \ac{ps}:
\begin{equation}
    \qop{X} = \qop{H} \cdot \qop{Z} \cdot \qop{H} \text{, and: }
    \qop{Y} = \qop{S} \cdot \qop{H} \cdot \qop{Z} \cdot \qop{H} \cdot \qop{S}^{\dag}
\end{equation}
As \ac{ps} are unitary matrices, then $ \ket{\perp} = \emptyset $.
Once in the diagonal basis, the repetition of the process describes in \Cref{subsec_qubitized_tensor_product} construct a 'ladder' of $\qop{CX}$ gates (dashed box part on the example \Cref{fig_qc_hs_ps}).
It is the basis change $\qop{V}_{i}$ that correspond to the dashed boxed part of \Cref{fig_qc_hs_ps}.
Note that the parity of the number of qubit that reads $\ket{1}$ (it Hamming weight) determine to which degenerated subspace a given state belong.
The previously mentionned 'ladder' of $\qop{CX}$ gates project this parity on a single qubit.

\subsection{\acl{scb}}
\begin{figure}[tb]
    \begin{center}
        {\includegraphics{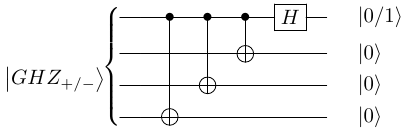}}
    \end{center}
    \caption{
    $4$ qubits \ac{ghz} basis transformation toward the computational basis which correspond to a projector-controlled-one-qubit-gate where the first qubit control the non-zero subspace rotation.
    }
    \label{fig_qc_ghz}
\end{figure}

For the \ac{scb} operator \ac{hs}, few usual circuit identity \cite[Section III]{ollive_gate_2025}\footnote{Note that in the author opinion, this new proof, is much straightforward than the \cite{ollive_gate_2025}'s one that is based on a separation of the three family of operators (\ac{ps} included) instead of a recursive combination based on \Cref{subsec_qubitized_tensor_product}.} are required:
\begin{itemize}
    \item The controlled gate derives from the $\qop{n}$ operator \ac{hs}:
    \begin{equation}
        \qop{P}(\theta) = e^{\ii \theta \qop{n}} \text{, and: }
        \qop{C e^{\ii \theta H_{p}}} = e^{\ii \theta \qop{n} \otimes \qop{H_{p}}}
    \end{equation}
    \item The eigenstates of a tensor product of ladder operators is a \ac{ghz} state of the same size:
    \begin{equation}
    \begin{aligned}
        \qop{\sigma}^{\otimes N} & + \qop{\hc} \\
        & = \ketbra{GHZ_{+}} - \ketbra{GHZ_{-}}
    \end{aligned}
    \end{equation}
    with the $N$ qubits \ac{ghz} state:
    \begin{equation}
        \ket{GHZ_{\pm}} = \frac{\ket{0}^{\otimes N} \pm \ket{1}^{\otimes N}}{\sqrt{2}}
    \end{equation}
    and the quantum circuit to construct \ac{ghz} state is well known and recalled in \Cref{fig_qc_ghz}.
    \item The basis change that transform the different ladder operator and number of exitation into each other are $\qop{X}$ gate:
    \begin{equation}
    \begin{aligned}
        \qop{n} = \qop{m}^{\ddag} \text{, and: } \qop{\sigma}^{\dag} = \qop{\sigma}^{\ddag} \text{, with: }
        \qop{A}^{\ddag} & = \qop{X} \cdot \qop{A} \cdot \qop{X}
    \end{aligned}
    \end{equation}
\end{itemize}

\begin{figure}[tb]
    \begin{center}
        \resizebox{\linewidth}{!}
        {\includegraphics{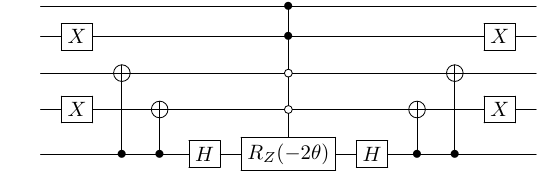}}
    \end{center}
    \caption{
    Decomposition of the unitary gate: $ e^{- \ii \theta \qop{n} \qop{m} \qop{\sigma} \qop{\sigma}^{\dag} \qop{\sigma} + \qop{\hc}} $.
    }
    \label{fig_qc_hs_scb}
\end{figure}

The two first items properties directly translate into arbitrary product of $\qop{n}$ and $\qop{\sigma} $ (plus the whole operator $ \qop{\hc}$) \ac{hs}, it is then possible to had $\qop{X}$ gates on both side to implement the desired \ac{scb} operator \ac{hs}\footnote{
    It is relevant to recall that it is possible to deal with the $\qop{C^{n} U}$ gates with either a linear gate overhead (plus an ancilla qubit) \cite[Sec.7: A, B and E]{barenco_elementary_1995}, \cite[p.182]{nielsen_quantum_2010}, \cite[SectionV.A]{ollive_gate_2025} or a polynomial overhead \cite{claudon_polylogarithmic-depth_2024}.
    \label{foottest}
}.
See for example \Cref{fig_qc_hs_scb}.

\subsection{Hadamard Gate} 
Hadamard Gate belongs to \ac{sth}.
By extension, the Hadamard transform is an interesting example that may be used to map denser problems.
The state transition associated with its change of basis is expressed as:
\begin{equation}
\begin{aligned}
    \qop{H} & = \frac{1}{\sqrt{2}}
    \begin{bmatrix}
        1 & 1 \\
        1 & -1
    \end{bmatrix} 
    & = \qop{R_{Y}}(\frac{\pi}{4}) \cdot \qop{Z} \cdot \qop{R_{Y}}(\frac{\pi}{4})^{\dag}
\end{aligned}
\end{equation}

\begin{figure}[tb]
    \begin{center}
        \resizebox{\linewidth}{!}
        {\includegraphics{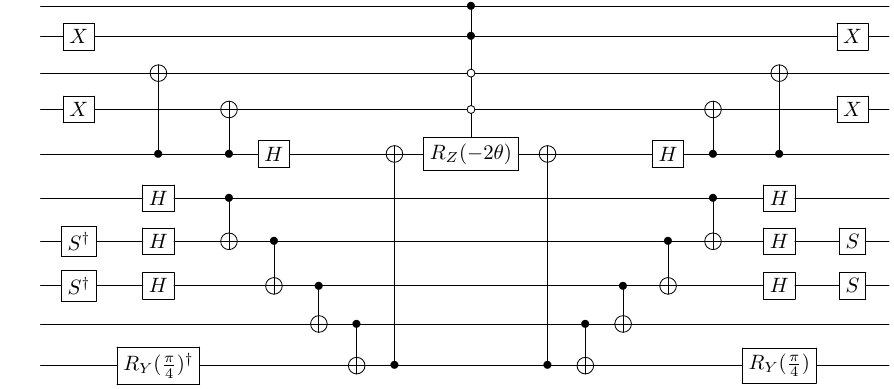}}
    \end{center}
    \caption{
    Decomposition of the unitary gate: $ e^{- \ii \theta \qop{n} \qop{m} \qop{\sigma} \qop{\sigma}^{\dag} \qop{\sigma} \qop{X} \qop{Y} \qop{Y} \qop{Z} \qop{H} + \qop{\hc}} $.
    }
    \label{fig_qc_hs_psscb}
\end{figure}

Again this three techniques along with any \ac{sth} can be combined by tensor product, for instance \Cref{fig_qc_hs_psscb}, thanks to \Cref{subsec_qubitized_tensor_product}.
The basic state transitions are gathered in \Cref{table_basis_change_simple}.

\begin{table}[tb]
  \caption{State transition required to diagonalize some single qubit \ac{sth}.}
  \center
    {\begin{tabular}{c|c|c|c|c|c}
      \textbf{$\qop{H_{i}}$} & $\qop{Z}$ & $\qop{X}$ & $\qop{Y}$ & $\qop{m}$ & $\qop{H}$ \\
      \hline
      \textbf{$\qop{V_{i}}$} & $\qop{I}$ & $\qop{H}$ & $ \qop{H} \cdot \qop{S}^{\dag} $ & $\qop{X}$ & $\qop{R_{Y}(\frac{\pi}{4})}^{\dag}$ \\
    \end{tabular}}
  \label{table_basis_change_simple}
\end{table}

\section{Well-Structured Matrices Decomposition in Special Tripotent Hamiltonian \label{sec_well_structured_problem}}
\cite[section V]{ollive_gate_2025} details how a mix of \ac{ps} and \ac{scb} allows for implementing problems that are naturally mapped with these operators.
It includes: second-quantized fermionic Hamiltonians, higher-order optimization problems, and component-by-component sparse matrices.
It also highlights that these operators can be used to implement structured matrices, namely the adjacency matrix associated with a graph: the regular Cartesian grid.
This section details how \ac{sth} can be used to implement structured matrices.
For this paper, we use the conventions $ n = 2^{N} \in \mathbb{N} $ and $ m = 2^{M} \in \mathbb{N} $. 
The studied matrices are implemented on a qubit register of size $M$ and thus (often) have size $m = 2^{M}$.

\subsection{Circulant and Toeplitz Matrix (Diagonals)} 
\begin{figure}[tb]
\begin{center}
    \resizebox{\linewidth}{!}
    {\includegraphics{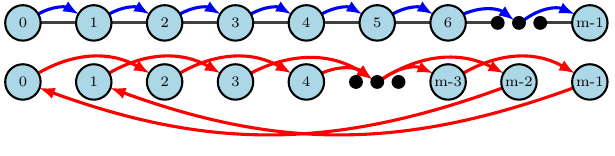}}
\end{center}
\caption{Symmetric Toeplitz matrix: $\qop{H}_{m-1, m}^{toep}$ in black, Toeplitz matrix: $\qop{M}(m - 1)$ in blue and directed circulant matrix: $\qop{M}(m - 2)$ in red.}
\label{fig_graph_circ}
\end{figure}

Toeplitz matrices $\qop{H}^{toep}$ are the matrices with constant values on each diagonal.
They can be written as the sum of their diagonals:
\begin{equation}
    \qop{H}^{toep} = \sum_{n = 1}^{m} \qop{H}_{n, m}^{toep}
\end{equation}
Each of these diagonal matrices can be represented as the graph adjacency matrix of a line of nodes that are all linked to their $(m - n)$\textsuperscript{th} neighbor and known as path graph \Cref{fig_graph_circ}.
We hereby detail how to decompose these matrices as a sum of \ac{sth}.
It is possible to \ac{be} non-Hermitian circulant using \Cref{section_non_hermitian} or the adder property \Cref{eq_circ_add_summand_decomp}.

\begin{figure}[tb]
    \hspace{-1.5 cm}
        \resizebox{10cm}{!}
        {\includegraphics{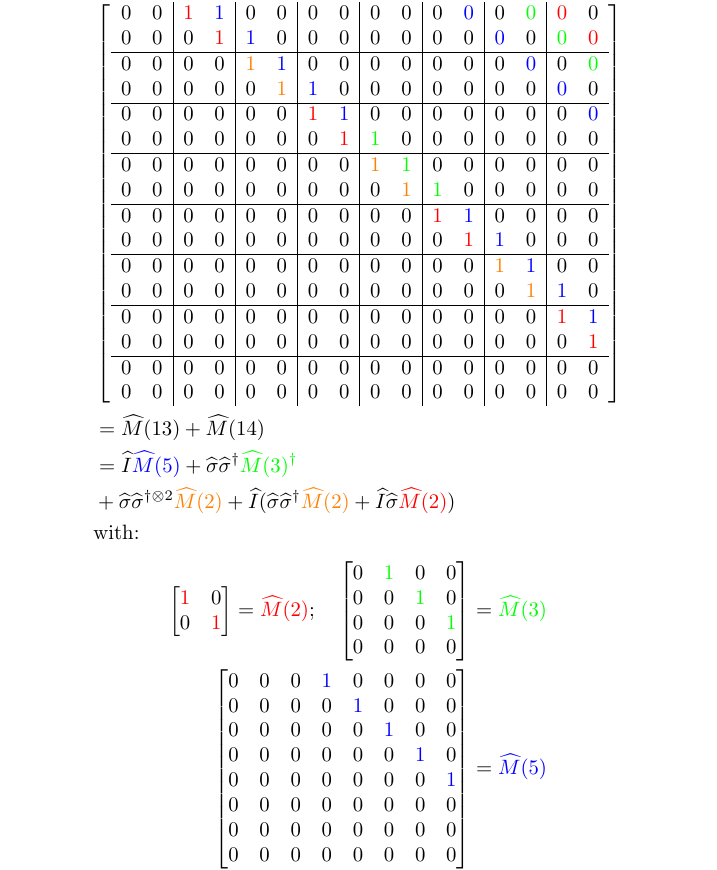}}
    \caption{Illustration of \Cref{eq_toep_summand_decomp_firststage}.}
    \label{fig_eq_circ}
\end{figure}

\paragraph{Toeplitz Matrices}
While \cite[Section V.C]{ollive_gate_2025} (and \cite[Fig. 1]{sato_hamiltonian_2024} indirectly) proposes a \ac{scb} plus \ac{ps} decomposition for the tridiagonal circulant (for which $ m = n - 1 $), it is possible to generalize it to any circulant (any $n$):
\begin{equation}
    \qop{H}_{n, m}^{toep} = \qop{\sigma}^{\otimes |M - w_{n}|} \qop{M}(n) + \qop{\hc}
    \label{eq_toep_summand_decomp}
\end{equation}
with $\qop{M}(n)$ defined by induction:
\begin{equation}
\begin{aligned}
    \qop{M}(n)
    & = \qop{I} \qop{\sigma}^{\otimes \alpha_{n}} \qop{M}(n - 2^{w_{n}}) + 
    \qop{\sigma} \qop{\sigma}^{\dag \otimes \beta_{n}} \qop{M}(2^{w_{n} + 1} - n)^{\dag} 
    \\
    & = \qop{I} \qop{A}(n) + \qop{\sigma} \qop{B}(n)
\end{aligned}
\label{eq_toep_summand_decomp_firststage}
\end{equation}
with the diagonal index $n$ (starting from the right of the first line),
\begin{equation}
\begin{aligned}
    n & \in [2^{w_{n}}, 2^{w_{n} + 1}[ \\
    \Leftrightarrow w_{n} & = \lfloor \log_{2}(n) \rfloor \\
    & = \alpha_{n} + \log_{2}(\mathrm{size}[\qop{M}(n - 2^{w_{n}})]) \\
    & = \beta_{n} + \log_{2}(\mathrm{size}[\qop{M}(2^{w_{n} + 1} - n)])
\end{aligned}
\end{equation}
and:
\begin{equation}
\begin{aligned}
    \qop{M}(2^{N}) & = \qop{I}^{\otimes N} \text{ , } N \neq 0 \\
    \qop{M}(1) & = \qop{\sigma} \\
    \qop{M}(3) & = \qop{\sigma} \qop{\sigma}^{\dag} +  \qop{I}\qop{\sigma} \\
\end{aligned}
\end{equation}
Note that: $ 2^{w_{n} + 1} - n = n - 2(n - 2^{w_{n}}) $.
This equation is illustrated by \Cref{fig_eq_circ}.

\paragraph{Circulant Matrices} They are specific Toeplitz matrices with two diagonals with the same value:
\begin{equation}
\begin{aligned}
    \qop{H}_{n, m}^{circ} = \qop{H}_{n, m}^{toep} + \qop{H}_{m - n, m}^{toep \dag}
    \label{eq_circ_summand_decomp}
\end{aligned}
\end{equation}
An illustrative quantum circuit is proposed in \Cref{{fig_qc_circ}}.
These matrices are the adjacency matrix of cycle graph.

\begin{figure*}[tb]
    \begin{center}
        \resizebox{\linewidth}{!}
        {\includegraphics{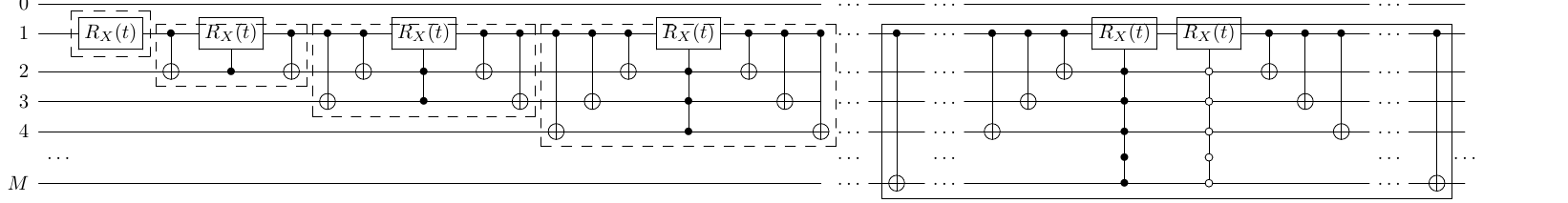}}
    \end{center}
    \caption{Quantum circuit associated with the \ac{hs} of the circulant: $ \qop{H}_{m - 2, m}^{circ} $ (only one Trotter repetition).
    Each summand's exact \ac{hs} is in a dotted box; the box contains the two pure \ac{scb} terms.}
    \label{fig_qc_circ}
\end{figure*}

\begin{figure}[tb]
    \begin{center}
        \resizebox{\linewidth}{!}
        {\includegraphics{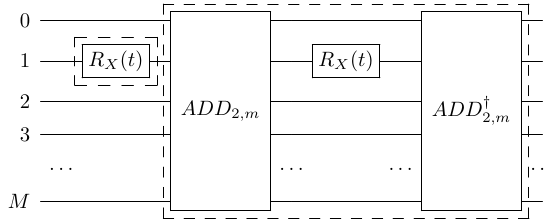}}
    \end{center}
    \caption{Quantum circuit associated with the adder-based variant to construct the \ac{hs} of the circulant: $ \qop{H}_{m - 2, m}^{circ} $ (only one Trotter repetition).
    Each summands exact \ac{hs} is in a dotted box.}
    \label{fig_qc_circ_add}
\end{figure}

\paragraph{Circulant Matrix, adder variant} To optimize the \ac{sth} decomposition (in a smaller number of summands and so needed arbitrary rotations), it is possible to use the adder variant when the terms can be gathered following \Cref{eq_circ_bis_summand_decomp}.
For connections with the $n$\textsuperscript{th} 
neighbor and $ m > 2^{w_{n} + 1} $ nodes:
\begin{equation}
\begin{aligned}
    & \qop{H}_{m - (2^{w_{n} + 1} - n), m}^{circ} 
    = \qop{I}^{\otimes (\varepsilon + 1)} \qop{A}(n) + \qop{\sigma}^{\dag \otimes (\varepsilon + 1)} \qop{B}(n) \\
    & \qquad + \sum_{l = 0}^{\varepsilon} \qop{I}^{\otimes \varepsilon - l} \qop{\sigma} \qop{\sigma}^{\dag \otimes l} \qop{B}(n) + \qop{\hc} \\
    & = \qop{I}^{\otimes (\varepsilon + 1)} \qop{A}(n) + \qop{I}^{\otimes \varepsilon} \qop{\sigma} \qop{B}(n) \\
    & \qquad + \qop{ADD_{2^{w_{n}}, m}}^{\dag} \cdot \qop{I}^{\otimes \varepsilon} \qop{\sigma} \qop{B}(n) \cdot \qop{ADD_{2^{w_{n}}, m}} + \qop{\hc}
    \label{eq_circ_bis_summand_decomp}
\end{aligned}
\end{equation}
with: $ 
    \varepsilon = M - 1 - \log_{2}(\mathrm{size}[\qop{B}(n)])$.
When $ n = 2^{N} $ is exactly a power of two, we instead have that:
\begin{equation}
\begin{aligned}
    \qop{H}_{m - n, m}^{circ} & = \qop{I}^{\otimes (M - N - 1)} \qop{\sigma} \qop{I}^{\otimes N}
    \\
    & + \qop{ADD_{n, m}}^{\dag} \cdot \qop{I}^{\otimes (M - N - 1)} \qop{\sigma} \qop{I}^{\otimes N} \cdot \qop{ADD_{n, m}} + \qop{\hc}
    \label{eq_adder_variant}
\end{aligned}
\end{equation}
This variant was proposed by \cite{ty_double-logarithmic_2024} to \ac{be} the tridiagonal circulant (for which $ n = m - 1 \Leftrightarrow N = 0 $).
As $M$ and $N$ are integers, this variant is efficient to implement diagonal corresponding to shifted Mersenne numbers $ 2^{M} - 2^{N} = 2^{N} (2^{M - N} - 1) $.
This adder variant quantum circuit, along with the associated reduction, is illustrated in the quantum circuits of \Cref{fig_qc_circ_add}.

The required change of basis for the second term is the \textbf{modular adder}:
\begin{equation}
    \qop{ADD_{n, m}} = \sum_{i = 0}^{m - 1} \ket{\binaire{(i + n)\mathrm{mod}[m]}} \bra{\binaire{i}}
\end{equation}
The modular adder is well studied, and two popular strategies for decomposing it into basic quantum gates are detailed in the appendix.
It is important to highlight that the modular adder matrix corresponds to the directed graph case, and so using \Cref{section_non_hermitian} allows us to construct the \ac{bei} of the adder.
This link works in both directions; the undirected circulant is also \ac{bed} using adders:
\begin{equation}
    \qop{H}_{n, m}^{circ} = \qop{ADD_{n, m}} + \qop{ADD_{n, m}}^{\dag}
    \label{eq_circ_add_summand_decomp}
\end{equation}
This strategy is used in several papers to \ac{be} tridiagonal (also refered to as banded) circulant matrices \cite{camps_explicit_2023, zecchi_block_2026}.
\cite[eq 13-16]{wan_block-encoding-based_2021} proposes an interesting variant of this method to \ac{be} Toeplitz from modified adders\footnote{Although this method falls outside the scope of the present study, its structural similarities allow it to be readily described using the paper’s notation, as detailed in the \Cref{appendix_adder}.}.
They also propose a variant adapter to construct Hankel matrices \cite[eq 24-25]{wan_block-encoding-based_2021}.

\subsection{Anti-Circulant and Hankel Matrix (Anti-Diagonal)}
\begin{figure}[tb]
\begin{center}
    \resizebox{\linewidth}{!}
    {\includegraphics{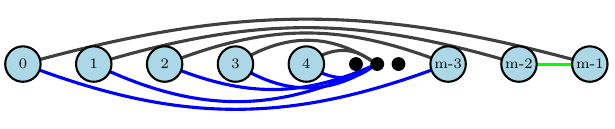}}
\end{center}
\caption{Main diagonal anti-circulant adjacency matrix: $\qop{H_{\frac{m}{2}, m}^{a-circ}}$ in black, Hankel matrix associated graph: $\qop{H}_{\frac{m}{2} - 1, m}^{hank}$ in blue and $\qop{H}_{2 m - 2, m}^{hank}$ in green.
The combination of the blue and green matrix is the anti-circulant: $\qop{H}_{\frac{m}{2} - 1, m}^{a-circ}$.}
\label{fig_graph_hank}
\end{figure}

Anti-Circulant and Hankel Matrices have constant value on each anti-diagonal.
These matrices can be obtained by the sum of each of these anti-diagonals:
\begin{equation}
 \qop{H}^{hank} = \sum_{n = 1}^{2 m - 1} \qop{H}_{n, m}^{hank}
\end{equation}
The graph corresponding to the $n$\textsuperscript{th} anti-diagonal matrix is the interaction between the $n$\textsuperscript{th} node with the first node plus the second node with the $n-1$\textsuperscript{th} node, etc as illustrated by \Cref{fig_graph_hank}.

\paragraph{Hankel Matrices} (anti-diagonal) are constructed using a strategy similar to the case of the Toeplitz matrices by interverting the gates $\qop{I}$ with $\qop{X}$, $\qop{\sigma}$ by $\qop{m}$, and $\qop{\sigma}^{\dag}$ by $\qop{n}$ when $ n > m $.
It leads to:
\begin{equation}
\begin{aligned}
    \qop{H}_{n, 2^{w_{n}}}^{hank} = \qop{N}(n) \qquad \qquad \qquad \qquad \qquad \qquad \qquad \qquad \\
\begin{aligned}
    & = \qop{X} \qop{m}^{\otimes \alpha_{n}} \qop{N}(n - 2^{w_{n}}) + 
    \qop{m} \qop{n}^{\otimes \beta_{n}} \qop{N}(2^{w_{n} + 1} - n)^{\ddagger} 
\end{aligned}
\end{aligned}
\end{equation}
with the diagonal index $ n \in [2^{w_{n}}, 2^{w_{n} + 1}] $ (starting from the left) 
and:
\begin{equation}
\begin{aligned}
    \qop{N}(2^{N}) & = \qop{X}^{\otimes N} \text{ , } N \neq 0 \\
    \qop{N}(1) & = \qop{m} \\
    \qop{N}(3) & = \qop{m} \qop{n} + \qop{X} \qop{m} \\
    \qop{N}^{\ddag} & = \qop{X}^{\otimes \log_{2}(\mathrm{size}[\qop{N}])} \cdot \qop{N} \cdot \qop{X}^{\otimes \log_{2}(\mathrm{size}[\qop{N}])} \\
\end{aligned}
\end{equation}
Then, this equation is extended for an arbitrary node number $ m > 2^{w_{n + 1}} $:
\begin{equation}
    \qop{H}_{n, m}^{hank} =
        \qop{m}^{\otimes |M - w_{n}|} \qop{N}(n) 
    \label{eq_hank_summand_decomp}
\end{equation}
When $ n > m $:
\begin{equation}
    \qop{H}_{n, m}^{hank} = \qop{H}_{2 m - n, m}^{hank \ddag}
\end{equation}

\paragraph{Anti-Circulant} As for the circulant from the Toeplitz matrices, it can be obtained thanks to a sum of two anti-circulant:
\begin{equation}
\begin{aligned}
    \qop{H}_{n, m}^{a-circ} = \qop{H}_{n, m}^{hank} + \qop{H}_{m + n, m}^{hank}
    \label{eq_acirc_summand_decomp}
\end{aligned}
\end{equation}
It is easier to construct it using the adder variant because the adders naturally transform the anti-circulant into another anti-circulant of the same parity (by shifting along the diagonal).
The anti-circulant that start on the $n^{\text{th}}$ anti-diagonal: 
\begin{equation}
\begin{aligned}
   \qop{H_{n, m}^{a-circ}} & = 
    \left\{
    \begin{array}{ll}
        \qop{ADD_{\frac{n}{2}, m}}^{\dag} \cdot \qop{H_{m, m}^{hank}} \cdot \qop{ADD_{\frac{n}{2}, m}} & \text{if } n \text{ even} \\
        \qop{ADD_{\frac{n-1}{2}, m}}^{\dag} \cdot \qop{H_{1, m}^{a-circ}} \cdot \qop{ADD_{\frac{n-1}{2}, m}} & \text{if } n \text{ uneven}
    \end{array}
    \right. \\
    \label{eq_acirc_add_summand_decomp}
\end{aligned}
\end{equation}

These anti-circulant matrices can also be understood as anti-adders, and be \ac{bed} as:
\begin{equation}
\begin{aligned}
    \qop{H_{n, m}^{a-circ}} & = \qop{X}^{\otimes M} \cdot \qop{ADD}_{n, m} \\
    & = \sum_{i = 0}^{m - 1} \ket{\binaire{(m - (i + n + 1))\mathrm{mod}[m]}} \bra{\binaire{i}}
\end{aligned}
\end{equation}
Even if anti-adders allows to shift matrices along the anti-diagonal, it cannot be used as a change of basis to construct a circulant with the sum of two Hermitian matrices by analogy with the anti-circulant.
It would require the anti-adders whose anti-symmetric are not their Hermitian conjugate, and so does not lead to a Hermitian matrix.

\subsection{Arbitrary Small Circular Permutation \label{subsec_permut}} 
\begin{figure}[tb]
\begin{center}
    {\includegraphics{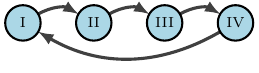}}
\end{center}
\caption{Circulant reordered by the permutation: $ \qop{U_{\mathrm{\pi}}} = \ket{I} \bra{\binaire{0}} + \ket{II} \bra{\binaire{1}} + \ket{III} \bra{\binaire{2}} + \ket{IV} \bra{\binaire{3}} $}
\label{fig_graph_rot}
\end{figure}

The graph representation of a circular permutation travels around all the nodes node by node, following the order of the computational basis reordered by the permutation.
It requires the ability to construct (an arbitrary) permutation:
\begin{equation}
    \qop{U_{\mathrm{\pi}}} = \sum_{i} \ket{\binaire{\mathrm{\pi}(i)}} \bra{\binaire{i}}
\end{equation}
with $\pi$ a permutation as illustrated with the graph \Cref{fig_graph_rot}.

This graph is the reordered circulant:
\begin{equation}
\begin{aligned}
    \qop{H}^{rot} & = \qop{Ur_{n, m}} + \qop{Ur_{n, m}}^{T} \\
    & = \qop{U_{\mathrm{\pi}}}^{\dag} \cdot \qop{ADD_{n, m}} \cdot \qop{U_{\mathrm{\pi}}} + \qop{U_{\mathrm{\pi}}}^{\dag} \cdot \qop{ADD_{n, m}}^{\dag} \cdot \qop{U_{\mathrm{\pi}}} \\
    & = \qop{U_{\mathrm{\pi}}}^{\dag} \cdot \qop{H}_{n, m}^{circ} \cdot \qop{U_{\mathrm{\pi}}}
    \label{eq_rot_summand_decomp}
\end{aligned}
\end{equation}
with $\qop{Ur_{n, m}}$ the directed graph adjacency matrix.
It is possible to \ac{be} directed graph associated to reordered circulant using \Cref{section_non_hermitian} or the adder property \Cref{eq_circ_add_summand_decomp}.

Note that:
\begin{itemize}
    \item the global permutation $\qop{U_{\mathrm{\pi}}}$ can also be the target matrix to implement when constructing the \ac{bei}.
    \item Depending on the permutation, it can be efficient to construct it \ac{hs} as a sparse matrix whose summands are pure \ac{scb} transition.
\end{itemize}

\paragraph{Construct the Unitary associated to Arbitrary Short-Cycle Permutations}
Arbitrary short-cycle permutations allow to construct directly small graph adjacency matrix.
The basis transition $\qop{U_{\mathrm{\pi}}}$ used to construct the $\qop{H}^{rot}$ change of basis is the permutation matrix:
\begin{equation}
    \qop{U_{\mathrm{\pi}}} = \sum_{i = 0}^{m - 1} \ket{\binaire{i}} \bra{\binaire{0}} \qop{Ur_{n, m}}^{(i + 1)}
\end{equation}
As any permutation can be decomposed 
as a product of reflection (transposition).
It requires as many reflections as the cycle length (minus one) $n_{\gamma} - 1$.

\textit{State-swapping} gates are used to implement very sparse matrices \cite[SectionV.D]{ollive_gate_2025}.
These gates are reflections or parity operators in the basis: $ \{ \frac{\ket{\lambda} \pm \ket{\lambda_{\perp}}}{\sqrt{2}} \} $.
These gates are constructed using only \ac{scb}, providing the possibility to swap two states that have a mutual eigenbasis:
\begin{equation}
\begin{aligned}
    \qop{C_{\ket{\perp}} X}\{ \ket{\lambda} ; \ket{\lambda_{\perp}} \}
    = \qop{\Pi}_{\ket{\perp}} + \ket{\lambda} \bra{\lambda_{\perp}} + \ket{\lambda_{\perp}} \bra{\lambda}
\end{aligned}
\end{equation}
Used in the computational basis, an $n-1$ end-to-end product of such gates allows for the implementation of arbitrary reordering of $n$ states:
\begin{equation}
\begin{aligned}
    \qop{U_{\mathrm{\pi}}} = \prod_{\ket{\gamma_{j}} \in R} \prod_{i = 0}^{n_{\gamma_{j}} - 2} 
    \qop{C_{\ket{\perp}} X} \{ \ket{\binaire{\sum_{k < j} (n_{\gamma_{k}} - 1) + i}} ; \qop{U_{\mathrm{\pi}}}^{(i + 1)} \ket{\gamma_{j}} \}
\end{aligned}
\end{equation}
with $ R = \{ \ket{\binaire{i}} \} / \{ \qop{U_{\mathrm{\pi}}}^{k} \ket{\binaire{i}} | 0 < k < n_{\gamma} ; k \in \mathbb{N} \} $, an element of each cycle orbit and $ n_{\gamma} $ the period of the associated circle: $ \ket{\gamma} = \qop{U_{\mathrm{\pi}}}^{n_{\gamma}} \ket{\gamma} $.

\paragraph{Construct the Unitary associated to Permutation Matrix Defined by an Explicit one-to-one Binary Function}
Graphs whose directed adjacency matrix corresponds to the permutation associated with an explicit one-to-one binary function can be \ac{be} on a quantum computer as follows:
\begin{equation}
    \qop{U_{\mathrm{\pi}}} = \qop{V_{\mathrm{\pi}}} \cdot \qop{SWAP}_{\mathcal{A} \leftrightarrow \mathcal{B}} \cdot \qop{V_{\mathrm{\pi}^{-1}}}
\end{equation}
with the unitaries\footnote{The convention used in \cite[subsection 3.2]{low_hamiltonian_2019} where matrices implemented in practice as unitary are expressed as isometries.
Typically, $ \qop{V_{\mathrm{\pi}}} = \sum_{x = 0}^{m - 1} \sum_{y = 0}^{m - 1}  \ketbra{\binaire{x}}_{\mathcal{A}} \otimes \ket{\binaire{y}} \bra{\binaire{\mathrm{f}(x, y)}}_{\mathcal{B}} $, but we are not interest in the output when $j \neq 0 $ as soon as it is compatible with the complete operator unitarity.
} $ \{ \qop{V_{\mathrm{\pi}}}; \qop{V_{\mathrm{\pi}^{-1}}} \} $ such that:
\begin{equation}
\begin{aligned}
    \qop{V_{\mathrm{\pi}}} & = \sum_{x = 0}^{m - 1} \ketbra{\binaire{x}}_{\mathcal{A}} \otimes \ket{0} \bra{\binaire{\mathrm{\pi}(x)}}_{\mathcal{B}} \\
    \qop{V_{\mathrm{\pi}^{-1}}} & = \sum_{x = 0}^{m - 1} \ketbra{\binaire{\mathrm{\pi}(x)}}_{\mathcal{A}} \otimes \ket{\binaire{x}} \bra{0}_{\mathcal{B}}
\end{aligned}
\end{equation}
It is worth noting that this process needs ancilla qubits, especially if the quantum arithmetic function: $ \{ \qop{V_{\mathrm{\pi}}}; \qop{V_{\mathrm{\pi}^{-1}}} \} $ construction induces a garbage register $\mathcal{C}$:
\begin{equation}
    \qop{V_{\mathrm{f}}} = \sum_{x = 0}^{m - 1} \ketbra{\binaire{x}}_{\mathcal{A}} \otimes \ket{0} \bra{\binaire{\mathrm{f}(x)}}_{\mathcal{B}} \otimes \ket{0} \bra{\beta}_{\mathcal{C}}
\end{equation}
Therefore, it must be cleaned using a reversible computing strategy following \cite[Table 1]{bennett_logical_1973}, \cite[Annex C, footnote 9]{ollive_quantum_2025}.
In other words, one-site quantum arithmetic can be constructed from general quantum arithmetic and used as a change of basis.

\subsection{Multiply by Two Matrix}
The multiply by $n$ matrix is used by \cite{camps_explicit_2023,sunderhauf_block-encoding_2024} (with $ n = 2 $) to implement other structured matrices (extended binary tree):
\begin{equation}
\begin{aligned}
    \qop{Mult}_{n, m} = & \ketbra{\binaire{m - 1}} + \\
    & \quad \sum_{i = 0}^{m - 2} \ket{\binaire{(i \times n)\modulo{m - 1}}} \bra{\binaire{i}}
\end{aligned}
\end{equation}
This matrix can easily be constructed as a unitary matrix for power of two:
\begin{equation}
    \qop{Mult}_{2^{n}, m} = \qop{SWAP}_{0, M-2} \cdot \prod_{i = 0}^{M - 1} \qop{SWAP}_{i+1, i}
    \label{eq_mult_by_two_unitary}
\end{equation}
It corresponds to a permutation of the bit indices.
It is interesting to notice that it is naturally decomposed by the \ac{scb} basis:
\begin{equation}
    \qop{Mult}_{2^{n}, m} = \qop{\sigma}^{\dag}_{0} \cdot \qop{\sigma}_{M-2} \cdot \sum_{i = 0}^{M - 1} \qop{\sigma}^{\dag}_{i + 1} \cdot \qop{\sigma}_{i}
    \label{eq_mult_by_two}
\end{equation}

\subsection{Matrices constructed as Outer Product}
\subsubsection{Pure State Density Matrix}
\begin{figure}[tb]
\begin{center}
    {\includegraphics{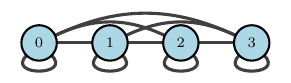}}
\end{center}
\caption{Graph associated with the adjacency matrix: $ 4 \qop{H}^{\otimes 2} \ketbra{00} \qop{H}^{\otimes 2} = 4 (\ketbra{+})^{\otimes 2} $.}
\label{fig_graph_density}
\end{figure}

The density matrix of a pure state:
\begin{equation}
    \ketbra{\psi} = \sum_{i} \sum_{j} \alpha_{i} \alpha_{j}^{*} \ket{\binaire{i}} \bra{\binaire{j}}
    \label{eq_density_summand_decomp}
\end{equation}
also known as the orthogonal projector, is \ac{bed} on a quantum computer.
If the component of $ \ket{\psi} = \sum_{i} \alpha_{i} \ket{\binaire{i}} $ are all real, the weighted graph associated with this Hermitian matrix is a dense graph that links all the nodes to all the others with different weight \Cref{fig_graph_density}.

The \ac{bei} is using \Cref{eq_projecteur_be}:
\begin{equation}
 \ketbra{\psi} = \qop{U_{\psi}} \ketbra{\binaire{0}} \qop{U_{\psi}}^{\dag}
\end{equation}
with: 
\begin{equation}
\begin{aligned}
    \qop{m}^{\otimes M} & = \ketbra{\binaire{0}}
    & = \frac{\qop{I} - \qop{X}^{\otimes M} \cdot \qop{C^{M-1} Z} \cdot \qop{X}^{\otimes M}}{2}
\end{aligned}
\end{equation}
and $ m = 2^{M} $ the number of components of $\ket{\psi}$ and $ \qop{U_{\psi}} = \ket{\psi} \bra{\binaire{0}} + \sum_{j = 1}^{\taille{\qop{U_{\psi}}} - 1} \ket{\psi_{\perp, j}} \bra{\binaire{j}} $.
As the construction of the state initialisation is linear in the number of components with different values, the implementing arbitrary $\qop{U_{\psi}}$ requires an exponential number of gates in the number of qubits.

It \ac{hs} naturally follows: 
\begin{equation}
\begin{aligned}
    e^{\ii t \ketbra{\psi}} = \qop{U_{\psi}} \cdot e^{\ii t \qop{m}^{\otimes M}} \cdot \qop{U_{\psi}}^{\dag} = (e^{\ii t} - 1) \ketbra{\psi} + \qop{I}
\end{aligned}
\end{equation}

\paragraph{Complete Graph}
A complete graph links all the nodes to all the vertices and it normalized adjacency matrix can easily be decomposed (up to a normalization factor) if its size $m$ is a power of two:
\begin{equation}
    \qop{K_{m}} = \ketbra{+}^{\otimes M} - \frac{1}{m} \qop{I}^{\otimes M}
    \label{eq_complete_graph}
\end{equation}

\subsubsection{Outer Product \acl{bei}} Their are also possible using a variant of the previous technique:
\begin{equation}
    \ket{\phi} \bra{\psi} = \qop{U_{\phi}} \ketbra{\binaire{0}} \qop{U_{\psi}}^{\dag}
    \label{eq_outer_summand_decomp}
\end{equation}
The Pseudo-Covariance case is a specific outer-product Matrix for which:
\begin{equation}
    \ket{\phi} = \ket{\psi}^{\star} 
    \Leftarrow \qop{U_{\phi}} = \qop{U_{\psi}}^{\star}
\end{equation}
Knowing $\qop{U_{\psi}}$ allows us to construct $\qop{U_{\phi}}$ with the reversed gate sequence.
These matrices can only be decomposed as \ac{lcu} to construct \ac{bei}. 
It does not provide directly the \ac{lch} to construct \ac{hs} or direct \ac{pvm}.
Interestingly, this result recovers \cite[Lemma 47]{gilyen_quantum_2019} in the case where the outer-product is a summand in a larger \ac{lcu}.

\paragraph{Line and column \ac{bei}} 
\begin{figure}[tb]
\begin{center}
    {\includegraphics{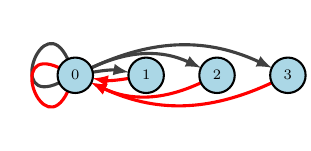}}
\end{center}
\caption{Graph associated with the adjacency matrix: $ 2 \qop{H}^{\otimes 2} \ketbra{00} $ in black and $ 2 \ketbra{00} \qop{H}^{\otimes 2} $ in red.}
\label{fig_graph_linecol}
\end{figure}
Line and column operators are constructed starting as specific case of outer-product matrix without one of the two basis transitions:
\begin{equation}
\begin{aligned}
    \ket{\psi} \bra{\binaire{j}} & = \qop{U_{\psi, j}} \ketbra{\binaire{j}} \\
    \ket{\binaire{j}} \bra{\psi} & = \ketbra{\binaire{j}} \qop{U_{\psi, j}}^{\dag}
    \label{eq_linecol_summand_decomp}
\end{aligned}
\end{equation}
with $ \ketbra{\binaire{j}} $ obtained by following \cite[SectionV.D]{ollive_gate_2025} and:
\begin{equation}
\begin{aligned}
    \qop{U_{\psi, j}} & = \ket{\psi} \bra{\binaire{j}} + \sum_{k = 0, k \neq j}^{\taille{\qop{U_{\psi}}} - 1} \ket{\psi_{\perp, k}} \bra{\binaire{k}} \\
    & = \qop{U_{\psi}} \underset{k \in \{i | \sum 2^{i} = j \} }{\otimes} \qop{X}_{k}
\end{aligned}
\end{equation}
The associated graph with a line or column is a directed star graph; all the nodes are linked toward or from a single node \Cref{fig_graph_linecol}.
An interesting case is the Hadamard transform: $ \qop{U_{\psi}} = \qop{H}^{\otimes M} $
that corresponds to a line or column with all the same value coefficients.

\paragraph{Star Graph}
The star graph normalized adjacency matrix whose size is a power of two is decomposed as:
\begin{equation}
    \qop{S_{m}} = (\ket{+} \bra{0})^{\otimes M} + (\ket{0} \bra{+})^{\otimes M} - \frac{2}{\sqrt{m}} \qop{m}^{\otimes M}
    \label{eq_star_graph}
\end{equation}
It is a Hermitian matrix with an easy outer product, but is difficult to decompose as an \ac{lch}.

\begin{figure}[tb]
\begin{center}
{\includegraphics{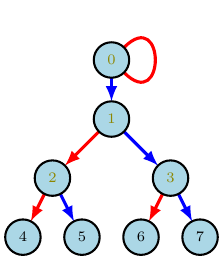}}
\end{center}
\caption{The perfect binary tree directed graph operators $ (\qop{I}^{\otimes 3} + \qop{ADD}_{1, 8}) \cdot \qop{Mult}_{2, 8} \cdot (\qop{I}^{\otimes 2} \otimes \ketbra{0}) $ with a direct identification of the operations: $\qop{I}^{\otimes 2} \otimes \ketbra{0}$ select only the $0$ of the most significant bit (green numbers), $\qop{Mult}_{2, 8}$ is the basic multiplication by two (in red), and $\qop{Mult}_{2, 8} \cdot \qop{ADD}_{1, 8}$ multiply by two and add one (in blue).}
\label{fig_graph_binary_tree}
\end{figure}

\paragraph{Perfect Binary tree}
A matrix proposed in the quantum computing bibliography is the perfect (or extended) binary tree matrix \cite{camps_explicit_2023,sunderhauf_block-encoding_2024}.
This matrix is constructed thanks to a sum of two outer products with an easy-to-implement basis change:
\begin{equation}
\begin{aligned}
    \qop{BT}_{m} & = (\qop{I}^{\otimes M} + \qop{ADD}_{1, m}) \cdot \qop{Mult}_{2, m} \\
    & \qquad \cdot (\qop{I}^{\otimes (M - 1)} \otimes \ketbra{0}) - \qop{m}^{\otimes M} + \qop{\hc}
    \label{eq_binary_tree}
\end{aligned}
\end{equation}
This decomposition is illustrated by \Cref{fig_graph_binary_tree} and is inspired by the quantum circuit proposed in \cite[fig. 4.13]{camps_explicit_2023}.
It requires four outer products plus a projector. 
It is a Hermitian matrix with an easy outer product, but is difficult to decompose as an \ac{lch}.

\paragraph{Compagnion Matrix}
Compagnion matrices are \underline{non}-Hermitian matrices written as the sum of a Toeplitz plus a column matrix, following:
\begin{equation}
    \qop{CM}_{m}(\ket{\psi}) = \qop{ADD}_{1, m}^{\dag} - \qop{\sigma}^{\otimes M} + \ket{\psi} \bra{\binaire{m-1}}
    \label{eq_compagnon_matrix}
\end{equation}
Using the previous definition of $ \ket{\psi} = \sum_{i} \alpha_{i} \ket{\binaire{i}} $, the characteristic polynomial of the matrix is: $ \sum_{i} \alpha_{i} x^{i} $.
Under certain diagonalizability conditions, these kinds of matrices can be used to find roots of high-degree polynomials.
Unfortunately, the implementation cost of the five required unitary matrices is quickly dominated by the cost of the state preparation to obtain $\ket{\psi}$, which severely restricts the set of polynomial coefficients that can be efficiently processed.

\subsection{Checkerboard Matrix}
Another matrix proposed in the quantum computing bibliography is the Checkerboard matrix \cite{sunderhauf_block-encoding_2024}.
Once normalized, this hermitian matrix is:
\begin{equation}
\begin{aligned}
    & \qop{CB}_{m} = \sum_{i = 0}^{m - 1} \sum_{j = 0}^{m - 1} \alpha_{i, j} \ket{\binaire{i}} \bra{\binaire{j}} \\
    & \text{ with: } \alpha_{i, j} = 
    \left\{
    \begin{array}{ll}
        0 & \text{if } i + j \text{ even} \\
        \frac{2}{m} & \text{if } i + j \text{ uneven}
    \end{array}
    \right. \\
\end{aligned}
\end{equation}
Such matrices are typically included in the \ac{sth} framework:
\begin{equation}
    \qop{CB}_{m} = (\ketbra{+})^{\otimes M - 1} \otimes \qop{X}
    \label{eq_checkerboard}
\end{equation}
It is also simple to construct the $90^{o}$ rotated board using:
\begin{equation}
\begin{aligned}
    & \qop{RCB}_{m} = \sum_{i = 0}^{m - 1} \sum_{j = 0}^{m - 1} \alpha_{i, j} \ket{\binaire{i}} \bra{\binaire{j}} \\
    & \qquad \quad = (\ketbra{+})^{\otimes M - 1} \otimes \qop{I} \\
    & \text{ with: } \alpha_{i, j} = 
    \left\{
    \begin{array}{ll}
        \frac{2}{m} & \text{if } i + j \text{ even} \\
        0 & \text{if } i + j \text{ uneven}
    \end{array}
    \right. \\
\end{aligned}
\end{equation}

\subsection{Matrices constructed as Graph Product}
\subsubsection{Tensor Product}
\begin{figure}[tb]
    \begin{center}
    \resizebox{\linewidth}{!}
    {\includegraphics{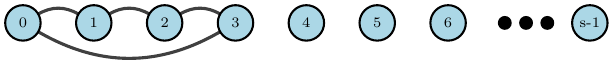}}
    \end{center}
    \caption{Graph whose adjacency matrix is $ \qop{H}_{3, 4, s}^{circ} $.}
    \label{fig_graph_bc}
\end{figure}

\paragraph{Block-Composition and Portion of Matrix}
Thanks to \ac{scb}, it is possible to compose a matrix as a sum of smaller \ac{bed} matrices.
It allows the construction of portions of diagonal, anti-diagonal, line, columns, \dots
The illustrative example \Cref{fig_graph_bc} encodes a circulant in the top-left corner of the matrix: 
\begin{equation}
    \qop{H}_{n, m, s}^{circ} = \qop{m}^{\otimes (S- M)} \qop{H}_{n, m}^{circ}
\end{equation}
with $ s = 2^{S} $ the full matrix size.
This kind of product between identity matrices, \ac{scb}, and Toeplitz or circulant matrices is well-suited to implementing matrices known as Block-Toeplitz-Toeplitz-Block (BTTB) and Block-Circulant-Circulant-Block (BCCB), which are used in fields such as signal processing and \ac{pde}.

\paragraph{Based on Complete Graph tensor product}
Commun named graph based on complete graph tensor product are $n$-crown graph: $ \qop{K_{2}} \otimes \qop{K_{n}} $, and $m \times n$ rook complement graph: $ \qop{K_{m}} \otimes \qop{K_{n}} $.

\subsubsection{Cartesian Product}
The adjacency matrix resulting from the cartesian product of two graphs is easily expressed from the tensor product of these graphs' adjacency matrices ($\qop{A_{G_{1}}}$ and $\qop{A_{G_{2}}}$) with the identity matrix:
\begin{equation}
\begin{aligned}
    \qop{A_{G}} & = \qop{A_{G_{1}}} \square \qop{A_{G_{2}}} \\
    & = \qop{A_{G_{1}}} \otimes \qop{I}^{\otimes \taille{G_{2}}} + \qop{I}^{\otimes \taille{G_{1}}} \otimes \qop{A_{G_{2}}}
\end{aligned}
\end{equation}

This section shows how the \ac{sth} framework benefits from it to implement interesting graphs that belong to the family of the graph Cartesian products, namely grid graphs, torus grid graphs, and Hamming graphs.
Other graphs of this family can be constructed as a cartesian product of the afore-mentioned graphs, which are listed in \Cref{appendix_graph_list}.
This list is constructed based on the list proposed in \cite{weisstein_graph_2026}.

\begin{figure}[tb]
\begin{center}
\resizebox{\linewidth}{!}
{\includegraphics{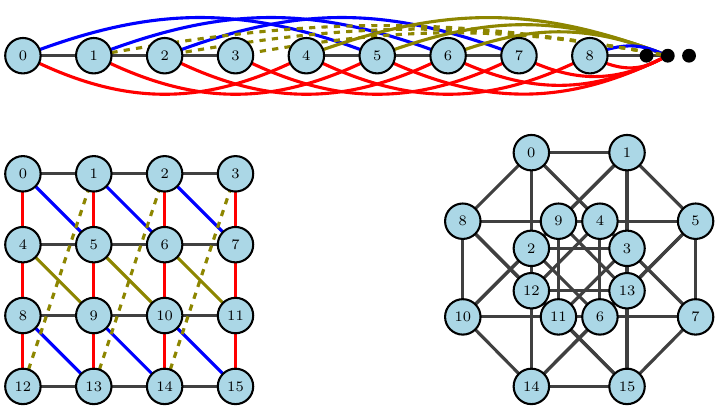}}
\end{center}
\caption{Two representations of: $ \qop{I}^{\otimes 2} \qop{H}_{3, 4}^{Toep} + \qop{H}_{3, 4}^{Toep} \qop{I}^{\otimes 2} + \qop{H}_{3, 8}^{Toep} \otimes \qop{I} + \qop{ADD_{4, 16}}^{\dag} \cdot (\qop{H}_{3, 8}^{Toep} \otimes \qop{I}) \cdot \qop{ADD_{4, 16}} $ (the first term is drawn in black, the second term is drawn in red, the their in blue and the fourth in green).
The three black dots represent the vertices from $9$ to $15$.
It also contains the Hamming graph $\qop{HG}_{2, 4}$ (lower-right figure).}
\label{fig_graph_grid}
\end{figure}

\paragraph{Cartesian Grid Graph}
It is possible to implement the adjacency matrix of a grid illustrated by \Cref{fig_graph_grid} as the sum of the adjacency matrices of all the dimension tensorial products with the identity matrix of the size of the other dimensions.
The tensorial product identity, applied to the other grid dimension, allows the pattern to be repeated in all dimensions.
As the number of vertices is multiplied by two for each new qubit, each new qubit allows for doubling the number of nodes in a dimension, whatever the size of the grid in the other dimension.
This explains why, for some structured lattices, the quantum algorithms are less subject to the curse of dimensionality \cite{jin_quantum_2023}.

We achieve two objectives at once by using \Cref{fig_graph_grid} to illustrate the implementation of the matrix diagonals with a Toeplitz matrix on the register encoding the two dimensions.
To change the Toeplitz matrices with circulant matrices would lead to cyclic boundary conditions (\textit{a.k.a.} torus grid graph).

\paragraph{Hamming Graph}
The Hamming graph is the graph Cartesian product of $d$ copies of the complete graph of size $ k = 2^{K} $:
\begin{equation}
\begin{aligned}
    \qop{HG}_{d, k} & = \qop{K_{k}}^{\square d} \\
    & = \sum_{i = 0}^{d - 1} \qop{I}^{\otimes (i K)} \otimes \qop{K_{k}} \otimes \qop{I}^{\otimes d - ((i + 1) K)}
    \label{eq_hamming_graph}
\end{aligned}
\end{equation}
These graphs are not only toy models but are also used in error-correction theory and in telecommunication networks.
The Hypercube graph: $ \qop{X}^{\square d} $, is the often used \ac{qaoa} mixing Hamiltonian \cite[eq.3]{farhi_quantum_2014}.
It is worth noting that all the summands of the Hamming graph commute, which allows us to do \ac{hs} without commutator error.
It can be used, for instance, to implement a continuous time \ac{qw} \cite{childs_exponential_2003} on a graph, such as \Cref{fig_graph_grid}.


\subsection{Summands Count} 
This section provides a quick overview of the cost of the previously mentioned structured matrix implementation using the proposed decomposition.
\Cref{table_cost} focuses on the number of terms (summands) of the Hermitian and unitary matrices decomposition, which is both related to the important \ac{ftqc} figure of merit: the number of arbitrary rotations and to the number of summands to sample when measuring the expectation value of the corresponding matrix.
The exact gate count depends on the chosen strategy to construct the multicontrolled gates used for the middle part of the circuit. 
This multicontrolled gate can be constructed with a linear cost in the number of qubits as detailled in \Cref{foottest}.
The number of summands is detailed for both the \ac{hs} and \ac{bei} of the problem matrix.
The number of summands is the same in both cases when the implemented matrix is a unitary matrix.

\begin{table}[tb]
  \caption{
    Maximum Number of Summands used to Decompose some Structured Matrices (with the associated \ac{hc}). \\ The Toeplitz, circulant, Hankel, and anti-circulant matrices' summands count are for a single diagonal or anti-diagonal\footnote{This distinction is quite important, as for a general displacement matrix, the number of diagonal entries is exponential in the number of qubits $M$, leading to an exponential number of summands unless there is a structure in the component values that can be exploited.} (plus \ac{hc}). \\
    Note that \Cref{eq_circ_add_summand_decomp} and \Cref{eq_mult_by_two_unitary} (marked by *) are two-summands decomposition.
    While these decompositions are not \ac{sth}- based, they are competitive in the number of summands for the \ac{lcu} case. \\
    Implementing circular permutation, density matrix, and outer product (marked by **) requires a basis change $\qop{V_{i}}$ constructed with an exponential number of gates for general instances.
    All the other decompositions require basis changes that are known and simple to construct. \\
    $ \mathrm{fusc}(n) $ is the Stern's diatomic sequence\footnote{The asymptotic behavior of this sequence is exponential in the worst case, but this exponential scales slowly on average as $ n^{\log_{2}(\frac{3}{2})} $, and logarithmically for values typically used in the state of the art.
    For instance $N$ gates for the Mersenne number $ n - 1 = 2^{N} - 1 $.}.
    }
  \center
  \resizebox{\linewidth}{!}
  {\begin{tabular}{|c|c|c|c|}
    \hline
    \textbf{Structured} & \textbf{Summands} & \multicolumn{2}{|c|}{\textbf{Summands Number}} \\
    \cline{3-4}
    \textbf{Matrix} & \textbf{Decomposition} & \textbf{\acs{lch}} & \textbf{\acs{lcu}} \\
    \hline
    Toeplitz & \Cref{eq_toep_summand_decomp} & $ \mathrm{fusc}(n) $ & $ 2 \mathrm{fusc}(n) $ \\
    \hline
    Circulant \underline{or} & \Cref{eq_circ_summand_decomp} & $ \mathrm{fusc}(n) +  $ & $ 2 ( \mathrm{fusc}(n) + $ \\
    Circular & \& \Cref{eq_circ_bis_summand_decomp} & $ \mathrm{fusc}(2^{w_{n} + 1} - n) $ & $ \mathrm{fusc}(2^{w_{n} + 1} - n) ) $ \\
    \cline{2-4}
    Permutation & \Cref{eq_adder_variant} if $ n = 2^{M} - 2^{N} $ & $2$ & $2$ \\
    \cline{2-4}
    ** & \Cref{eq_circ_add_summand_decomp}* & / & $2$ \\
    \hline
    Hankel & \Cref{eq_hank_summand_decomp} & $ \mathrm{fusc}(n) $ & $ 2 \mathrm{fusc}(n) $ \\
    \hline
    & \Cref{eq_acirc_summand_decomp} & $ \mathrm{fusc}(n) +  $ & $ 2 ( \mathrm{fusc}(n) + $ \\
    Anti-circulant & & $ \mathrm{fusc}(2^{w_{n} + 1} - n) $ & $ \mathrm{fusc}(2^{w_{n} + 1} - n) ) $ \\
    \cline{2-4}
    & \Cref{eq_acirc_add_summand_decomp} & $2$ & $2$  \\
    \hline
    Multiply & \Cref{eq_mult_by_two} & $M$ & $2 M$ \\
    \cline{2-4}
    by two & \Cref{eq_mult_by_two_unitary}* & / & $2$ \\
    \hline
    Density Matrix** & \Cref{eq_density_summand_decomp} & $1$ & $2$ \\
    \hline
    Complete Graph & \Cref{eq_complete_graph} & $2$ & $3$ \\
    \hline
    Outer Product** & \Cref{eq_outer_summand_decomp} & / & $2$ \\
    \hline
    Line \underline{or} Column & \Cref{eq_linecol_summand_decomp} & / & $2$ \\
    \hline
    Binary Tree & \Cref{eq_binary_tree} & / & $10$ \\
    \hline
    Checkerboard & \Cref{eq_checkerboard} & $1$ & $2$ \\
    \hline
    Hamming Graph & \Cref{eq_hamming_graph} & $2 d$ & $3 d$ \\
    \hline
  \end{tabular}}
  \label{table_cost}
\end{table}

\Cref{table_basis_change_cost} details the cost of the gates used to construct the change of basis associated with these structured matrices.
This basis change is polynomial except for the decomposition marked by **.
For decomposition marked by **, the basis change cost depends on the instance.

\begin{table}[tb]
  \caption{Implementation Cost of the non-trivial change of basis used to Construct the Structured Matrices.}
  \center
    {\begin{tabular}{|c|c|c|}
      \hline
      \textbf{State} & \multicolumn{2}{|c|}{\textbf{Implementation Cost}} \\
      \cline{2-3}
      \textbf{Transition} & \textbf{Count} & \textbf{Gate} \\
      \hline
      $\qop{ADD_{n, m}}$ & $2$ & $\qop{QFT}$ ($ \mathrm{O}(M^{2}) $) \\
      if $ n = 2^{N} $ & $M$ & $\qop{P}(\theta)$ \\
      \cline{2-3}
      and $ m = 2^{M} $ & $M$ & $\qop{C^{\otimes n} X}$ with $ n  \in \{ 1, M \} $ \\
      \hline
      & $m$ & $ \qop{C_{\ket{\perp}} X} \{ \ket{\binaire{i}} ; \ket{\binaire{j}} \}$ \\
      \cline{2-3}
      $\qop{U_{\mathrm{\pi}}}$ & $M$ & $\qop{SWAP}$ \\
      & $2$ & \ac{qa} \\
      \hline
      $\qop{U_{\psi}} = \qop{PREP} $ & $\mathrm{O}(m)$ & Gates \cite{grover_creating_2002, mottonen_quantum_2004,sun_asymptotically_2023} \\
      \hline
    \end{tabular}}
  \label{table_basis_change_cost}
\end{table}

Note that complete query construction is not studied, as it requires many implementation choices that are not fixed by our framework.
For \ac{hs} by \ac{pf}, Trotter repetition count is impacted by the \ac{pf} numerical error linked to the commutator structure. 
For \ac{bei} by \ac{lcu}, a practical implementation would involve a study to determine the most suitable strategy to factorize as many commuting summands basis change as possible.
All these external parameters would imply an in-depth study that is beyond the scope of this paper to determine which summand decomposition is, in practice, the most efficient.
Without addressing this point, what the \Cref{table_cost} and \Cref{table_basis_change_cost} teach us is that the proposed matrices can be implemented in a polynomial number of basic gates in the number of qubits when one of the listed strategies is used (apart from specific instances of the matrices marked by **).

\section{Conclusion}
This work completes previous work \cite{ollive_gate_2025} in two ways.
It provides a more general framework, called \ac{sth}, of which \ac{scb} plus \ac{ps} Hamiltonians are specific instances, as well as many other matrices.
This framework provides an intuitive understanding of this \ac{sth} through single-qubit identities.
It shows that the matrices constructed using this framework are compatible with the three main quantum computing queries: \ac{hs}, \ac{bei}, and cost functions based on \ac{pvm}.
This framework allows for simplifying the \ac{bei} qubitization procedure. 
It also completes \cite{ollive_gate_2025} by giving a list of structured matrices that cannot be directly mapped to \ac{ps} plus \ac{scb} but can efficiently be decomposed and implemented as a \ac{sth}.
These matrices can then be combined to construct, for instance, lattices on which stochastic problems, such as a \ac{qw} of discretized \ac{pde}, can be expressed.

This paper does not address the routines that combine the summands (\ac{pf} and \ac{bei} by \ac{lcu}) but proposes a decomposition of summands for structured matrices suited for problem encoding.
The intense expressivity allowed by the \ac{sth}, combined with the efficiency with which it allows to construct the three main quantum computing queries, makes this family of matrices particularly well-suited to encode problems to be addressed with quantum computers.
We are looking forward to see if further development on \ac{sth} will allow to decompose other structured matrices efficiently or to refine the bounds we can derive for queries constructed with these operators.

\bibliography{article_5_structured_matrices}

\section{Funding}
This research work was supported in part by the French PEPR integrated project Etude de la PIle Quantique — EPIQ, (ANR-22-PETQ-0007).

\section{Acknowledgement}
The authors thank Pierre-Emmanuel CLET for correcting some equations and for his help in improving the article's notations.

\appendix

\section{Explicit Example}
\paragraph{Circular Permutation by State-Swapping}
This sub-appendix illustrates how to implement a permutation that allows for the reordering of a circulant with an illustrative example:
\begin{equation}
\begin{aligned}
    \qop{Ur_{1, 4}} & = \qop{U_{\mathrm{\pi}}}^{\dag} \cdot \qop{ADD_{1, 4}} \cdot \qop{U_{\mathrm{\pi}}} \\
    & = \begin{bmatrix}
    0 & 0 & 0 & 1 \\
    0 & 1 & 0 & 0 \\
    1 & 0 & 0 & 0 \\
    0 & 0 & 1 & 0
    \end{bmatrix} \cdot
    \begin{bmatrix}
    0 & 0 & 0 & 1 \\
    1 & 0 & 0 & 0 \\
    0 & 1 & 0 & 0 \\
    0 & 0 & 1 & 0
    \end{bmatrix} \cdot
    \begin{bmatrix}
    0 & 0 & 1 & 0 \\
    0 & 1 & 0 & 0 \\
    0 & 0 & 0 & 1 \\
    1 & 0 & 0 & 0
    \end{bmatrix} \\
    & = \begin{bmatrix}
    0 & 0 & 0 & 1 \\
    0 & 0 & 1 & 0 \\
    1 & 0 & 0 & 0 \\
    0 & 1 & 0 & 0
    \end{bmatrix} \\
    & = \ket{\mathrm{bin}[2]} \bra{\mathrm{bin}[0]} + \ket{\mathrm{bin}[3]} \bra{\mathrm{bin}[1]} \\
    & + \quad \ket{\mathrm{bin}[1]} \bra{\mathrm{bin}[2]} + \ket{\mathrm{bin}[0]} \bra{\mathrm{bin}[3]}
\end{aligned}
\end{equation}
which means:
\begin{equation}
\begin{aligned}
    \qop{U_{\mathrm{\pi}}}^{\dag} & = \sum_{i = 0}^{3} \qop{Ur_{1, 4}}^{(i + 1)} \ket{\mathrm{bin}[0]} \bra{\mathrm{bin}[i]} \\
    & = \ket{\mathrm{bin}[2]} \bra{\mathrm{bin}[0]} + \ketbra{\mathrm{bin}[1]} \\
    & \qquad + \ket{\mathrm{bin}[3]} \bra{\mathrm{bin}[2]} + \ket{\mathrm{bin}[0]} \bra{\mathrm{bin}[3]}
\end{aligned}
\end{equation}
which is constructed using:
\begin{equation}
\begin{aligned}
    \qop{U_{\mathrm{\pi}}}^{\dag} & = \qop{C_{\ket{\perp}} X}\{ \qop{U_{\mathrm{\pi}}}^{2} \ket{\mathrm{bin}[0]} ; \ket{\mathrm{bin}[0]} \} \cdot \\
    & \qquad \qop{C_{\ket{\perp}} X}\{ \qop{U_{\mathrm{\pi}}} \ket{\mathrm{bin}[0]} ; \ket{\mathrm{bin}[0]} \} \\
    & = (\ket{\mathrm{bin}[3]} \bra{\mathrm{bin}[0]} + \ket{\mathrm{bin}[0]} \bra{\mathrm{bin}[3]} + \ketbra{\mathrm{bin}[1]} \\
    & \qquad + \ketbra{\mathrm{bin}[2]}) (\ket{\mathrm{bin}[2]} \bra{\mathrm{bin}[0]} + \ket{\mathrm{bin}[0]} \bra{\mathrm{bin}[2]} \\
    & \qquad + \ketbra{\mathrm{bin}[1]} + \ketbra{\mathrm{bin}[3]}) \\
    & = \ket{\mathrm{bin}[2]} \bra{\mathrm{bin}[0]} + \ketbra{\mathrm{bin}[1]} \\
    & \qquad + \ket{\mathrm{bin}[3]} \bra{\mathrm{bin}[2]} + \ket{\mathrm{bin}[0]} \bra{\mathrm{bin}[3]}
\end{aligned}
\end{equation}

\section{Adder Gate Decomposition \label{appendix_adder}}

\begin{figure}[tb]
\begin{center}
\resizebox{\linewidth}{!}{\includegraphics{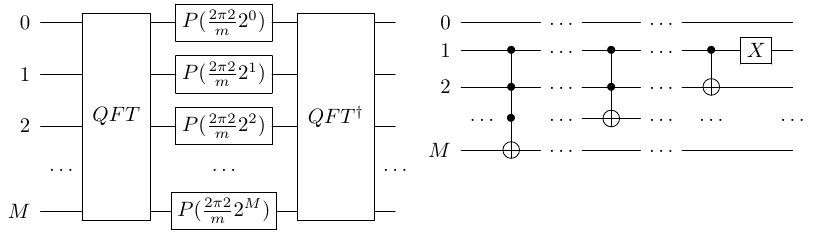}}
\end{center}
\caption{The two strategies to construct the quantum adder: $\qop{ADD}_{2, m}$.}
\label{fig_qc_adders}
\end{figure}

\paragraph{Adder Spectral Decomposition}
The \ac{hs} at specific times of the Hamiltonian circulant matrix:
\begin{equation}
\begin{aligned}
    \qop{H_{add}} & = \qop{QFT}^{\dag} \cdot (\sum_{i = 0}^{m - 1} i \ketbra{\mathrm{bin}[i]}) \cdot \qop{QFT} \\
    & = \sum_{k = 0}^{m - 1} c_{k} \qop{ADD}_{k, m} \\
    \text{with } & c_{k} = \frac{1}{m} \sum_{j = 0}^{m - 1} \lambda_{j} e^{i 2 \pi \frac{j k}{m}} 
    \text{ and } \qop{ADD}_{0, m} = \qop{I}^{\otimes M} \\
    \text{or } & \lambda_{j} = j \text{ and } c_{k}^{\dag} = c_{m - k}
\end{aligned}
\end{equation}
allows to \ac{be} any circulant matrix:
\begin{equation}
\begin{aligned}
    \qop{U_{add}}(t) & = e^{- \ii t \qop{H_{a}}} \\ 
    & = \sum_{k = 0}^{m - 1} c_{k} \qop{ADD}_{k, m} \\
    \text{with } & c_{k} = \frac{1}{m} \sum_{j = 0}^{m - 1} e^{\ii (2 \pi\frac{j k}{m} - t j)} 
\end{aligned}
\end{equation}
more specifically, any adder-like circular permutation:
\begin{equation}
    \qop{U_{add}}(t = \frac{2 \pi n}{m}) = \qop{ADD}_{n, m}
\end{equation}
As the diagonal matrix is constructed with a phase gate on each qubit, the construction cost of this unitary matrix is mainly the size $M$ $\qop{QFT}$, which is quadratic with respect to $M$.
Note that the basis transition $\qop{QFT}$ also allows for sampling the expectation value of this matrix by \ac{pvm}.
An example of such implementation can be found in \Cref{fig_qc_adders}.

These results are derived by a different approach in the context of the Quantum Arithmetic-based adders \cite{draper_addition_2000}, \cite[section 4]{ruiz-perez_quantum_2017}, \cite[Section 3.1.1]{ollive_quantum_2025} and with a similar approach in \cite[eq 7]{wan_asymptotic_2018}.

\paragraph{One-Site Quantum Arithmetic-based Adder}
Another usual strategy to construct the adder is to use one-site quantum arithmetics:
\begin{equation}
    \qop{ADD}_{n, m} = \prod_{i = N + 1}^{M - 1} \qop{C^{M - i - 1} X_{i}} \text{ with } n = 2^{N}
\end{equation}
For adding values that are not powers of two, it is required to use the property:
\begin{equation}
\begin{aligned}
    \qop{ADD}_{a + b, m} & = \qop{ADD}_{a, m} \cdot \qop{ADD}_{b, m} \\
    \qop{ADD}_{a - b, m} & = \qop{ADD}_{a, m} \cdot \qop{ADD}_{b, m}^{\dag}
    \label{eq_adder_aditivity}
\end{aligned}
\end{equation}
with $ a, b \leq m $.
An example of such implementation can be found in \Cref{fig_qc_adders}.

One-site arithmetic also allows us to construct other permutations that can be relevant to the \ac{bed} problem and may be used as a change of basis to construct \ac{hs}.
This strategy is used in several paper to \ac{be} banded circulant matrices \cite{camps_explicit_2023, zecchi_block_2026}.
An example is the product-by-$2$ gate \cite[Figure 9]{camps_explicit_2023}.

\paragraph{Toeplitz based on Modified Adders}
\cite[eq 13]{wan_block-encoding-based_2021} proposed an alternative to construct Toeplitz matrices with a \ac{lcu} based on modified adders:
\begin{equation}
\begin{aligned}
   \qop{Toep}_{n, m} & = \qop{\sigma}^{\otimes |M - w|} \qop{M}(n) \\
   & = \frac{\qop{ADD}_{m - n, m} + \qop{Z_{ADD}}_{m - n, m}}{2}
\end{aligned}
\end{equation}
with:
\begin{equation}
\begin{aligned}
   \qop{Z_{ADD}}_{n, m} & = \qop{Z_{-1, m}}^{n} \\
   \qop{Z_{f, 4}} & = 
   \begin{bmatrix}
    0 & 0 & 0 & f \\
    1 & 0 & 0 & 0 \\
    0 & 1 & 0 & 0 \\
    0 & 0 & 1 & 0
    \end{bmatrix}
\end{aligned}
\end{equation}
Note that these gates respect \Cref{eq_adder_aditivity} and can be implemented using either:
\begin{equation}
\begin{aligned}
   \qop{Z_{ADD}}_{n, m} & = \qop{ADD}_{n, m} \cdot \qop{C^{M - N} I^{\otimes N-1} (- I)} \\
   \text{ with } n & = 2^{N} 
   \text{ and } - \qop{I} = \qop{Z} \cdot \qop{X} \cdot \qop{Z} \cdot \qop{X}
\end{aligned}
\end{equation}
or using the operator of the \cite{childs_quantum_2017}'s \ac{lcu} circuit.
\cite{wan_block-encoding-based_2021}'s method is easily extendable to Hankel matrices \cite[eq 24-25]{wan_block-encoding-based_2021}.

\section{Computable Bounds}
The properties highlighted in this appendix are not specific to \ac{sth}, but follow for any decomposition with terms whose spectral norm is at most one.
The advantage is that this is directly true without extra renormalization for \ac{sth} operators.

\subsection{Statistical Sampling Variance}
Quickly converging expectation values measured by statistical methods are a quantum information challenge for \ac{vqa} \cite[section 4]{mcclean_theory_2016}.
It is known that the variance of each \ac{pvm} of the problem Hamiltonian summands scale as the inverse of the number of shots (and so the standard deviation as the inverse square root).
The prefactor attached to the variance is unknown in the general case \cite[eq 60]{mcclean_theory_2016}. 
This section emphasizes that the variance of each \ac{pvm} of the problem Hamiltonian summands can be bounded if the Hamiltonian is only decomposed as a sum of \ac{sth}.
For each \ac{sth}, once the basis transition toward the computational basis is done, the measurement of the expectation value is reduced to evaluating a single qubit or the parity (hamming weight) of a quantum register.
The projector's expectation value statistical variance can be bounded to:
\begin{equation}
\begin{aligned}
    \mathrm{Cov}[\qop{H_{i}}, \qop{H_{j}}, \ket{\psi}] & = \bra{\psi} \qop{H_{i}} \qop{H_{j}} \ket{\psi} - \bra{\psi} \qop{H_{i}} \ket{\psi} \bra{\psi} \qop{H_{j}} \ket{\psi} \\
    \mathrm{Var}[\qop{H_{i}}, \ket{\psi}] & = \mathrm{Cov}[\qop{H_{i}}, \qop{H_{i}}, \ket{\psi}] \\
    & \in [0, 1]
\end{aligned}
\end{equation}
as:
\begin{equation}
\begin{aligned}
    0 \leq |\bra{\psi} \qop{H_{i}} \ket{\psi}|^{2} & \leq \bra{\psi} \qop{H_{i}}^{2} \ket{\psi} \leq 1 \\
    |\mathrm{Cov}[\qop{H_{i}}, \qop{H_{j}}, \ket{\psi}]| & \leq 1
\end{aligned}
\end{equation}
as: $|\mathrm{Cov}[\qop{H_{i}}, \qop{H_{j}}, \ket{\psi}]|^{2} \leq \mathrm{Var}[\qop{H_{i}}, \ket{\psi}] \mathrm{Var}[\qop{H_{j}}, \ket{\psi}]$.

The independently sampled \ac{sth} expectation variances are then summed to obtain the variance of the problem Hamiltonian estimator $\overset{\sim}{H_{p}}$ \cite[eq 58-59]{mcclean_theory_2016}, \cite[eq 5]{crawford_efficient_2021}:
\begin{equation}
\begin{aligned}
    \bra{\psi} \qop{H_{p}} \ket{\psi} & = \sum_{i} \alpha_{i} \bra{\psi} \qop{H_{i}} \ket{\psi} \\
    \mathrm{Var}[\overset{\sim}{H_{p}}, \ket{\psi}] & = \sum_{i} \frac{\alpha_{i}^{2}}{N_{i}} \mathrm{Var}[\qop{H_{i}}, \ket{\psi}] 
    & \leq \sum_{i} \frac{|\alpha_{i}^{2}|}{N_{i}}
\end{aligned}
\label{eq_variance}
\end{equation}
With $N_{i}$ the number of shots allocated to the $i^{\text{th}}$ summand evaluation.
When the same basis transition allows to measure several \ac{sth} simultaneously (in the $g$ groups $G_{k}$), the variance becomes \cite[similarly to eq 20]{crawford_efficient_2021}:
\begin{equation}
\begin{aligned}
    \mathrm{Var}[\overset{\sim}{H_{p}}, \ket{\psi}] & = \sum_{k = 0}^{g} \sum_{i, j \in G_{k}} \frac{\alpha_{i} \alpha_{j}}{N_{k}} \mathrm{Cov}[\qop{H_{i}}, \qop{H_{j}}, \ket{\psi}] \\
    & \leq \sum_{k = 0}^{g} \sum_{i, j \in G_{k}} \frac{|\alpha_{i} \alpha_{j}|}{N_{k}} = \sum_{k = 0}^{g} \frac{1}{N_{k}} (\sum_{i \in G_{k}} |\alpha_{i}|)^{2}
\end{aligned}
\label{eq_variance_biss}
\end{equation}
With $N_{k}$ the number of shots allocated to the $k^{\text{th}}$ group evaluation.

\subsection{\acl{pf} Commutator Error}
Quickly converging \ac{pf} error is challenging for quantum algorithm based on the \ac{hs} query \cite{trotter_product_1958, childs_theory_2021}.
This error is typically expressed as a sum of nested commutators of the problem Hamiltonian summands \cite[eq.A18]{mehendale_estimating_2025}, \cite[Appendix.B]{ronfaut_numerical_2026}, \cite{kaoru_commutator_2026}.
An illustrative example is the Lie-Trotter formula's first order error:
\begin{equation}
\begin{aligned}
    (\prod_{i} e^{- \ii \frac{t}{n} \qop{\alpha_{i} H_{i}}})^{n} & = e^{- \ii t \qop{H} + \frac{t^{2}}{n} \qop{\Xi_{1}} + \mathcal{O}(\frac{t^{3}}{n^{2}})} \\
    \qop{\Xi_{1}} & = \frac{1}{2} \sum_{j < k} [\alpha_{j} \qop{H_{j}}, \alpha_{k} \qop{H_{k}}]
\end{aligned}
\end{equation}
As the spectral norm is both an upper bound of the Hamiltonian expectation value and a sub-multiplicative norm, it is possible to compute the upper bound of the commutation error if the problem Hamiltonian is decomposed with \ac{sth}:
\begin{equation}
\begin{aligned}
    |\bra{\psi} \qop{H} \ket{\psi}| & \leq ||\qop{H}|| \\
    ||\qop{H_{i}} \qop{H_{j}}|| & \leq ||\qop{H_{i}}|| ||\qop{H_{j}}|| \\
    ||[\alpha_{j} \qop{H_{j}}, \alpha_{k} \qop{H_{k}}]|| & \leq 2 | \alpha_{j} \alpha_{k} | \\
    || [\alpha_{i} \qop{H_{i}}, [\alpha_{j} \qop{H_{j}}, \alpha_{k} \qop{H_{k}}]] || & \leq 4 | \alpha_{i} \alpha_{j} \alpha_{k} | \\
    \text{and so: } ||\qop{\Xi_{1}}|| & \leq \sum_{j < k} | \alpha_{j} \alpha_{k} |
\end{aligned}
\end{equation}
It gives an easy-to-compute upper bound for the speed of convergence with respect to the number of repetitions (Trotter number) $n$.

\section{Circulant Decomposition and Stern's Diatomic Sequence}
This appendix explains why the tensorial product of $2 \times 2$ matrices decompositions of the $n$ diagonal needs as many summands as Stern's diatomic sequence $ \mathrm{fusc}(n) $.
Our induction to decompose the diagonal can be expressed as:
\begin{equation}
    \left\{
    \begin{array}{ll}
        w & = \lfloor \log_{2}(n) \rfloor \\
        p & = n - 2^{w_{n}} ; q = 2^{w_{n} + 1} - n \\
        \mathrm{f}(n) & = \mathrm{f}(p) + \mathrm{f}(q)
    \end{array}
    \right.
\end{equation}
with $\mathrm{f}(n)$ the number of summands arising when \Cref{eq_toep_summand_decomp} is used.
Stern's diatomic sequence is expressed as:
\begin{equation}
    \left\{
    \begin{array}{ll}
        \mathrm{fusc}(0) & = 0 ; \mathrm{fusc}(1) = 1 \\
        \mathrm{fusc}(2 k) & = \mathrm{fusc}(k) \\
        \mathrm{fusc}(2 k + 1) & = \mathrm{fusc}(k + 1) + \mathrm{fusc}(k)
    \end{array}
    \right.
\end{equation}
When $n$ even: $ n \leftarrow 2 n \Rightarrow w \leftarrow 1 + \lfloor \log_{2}(n) \rfloor $ and so:
\begin{equation}
    \mathrm{f}(2 n) = \mathrm{f}(2 p) + \mathrm{f}(2 q)
\end{equation}
this linearity implies that the factor two is reported on the next step of the induction.
At the last step, it simply changes the decomposition to twice the power of two while not changing the number of summands:
\begin{equation}
\begin{aligned}
    \mathrm{f}(n) & = \sum_{i = 0} \alpha_{i} \mathrm{f}(2^{i}) \text{ with } \alpha_{i} \in \mathbb{N} \\
    \Rightarrow \mathrm{f}(2 n) & = \sum_{i = 0} \alpha_{i} \mathrm{f}(2^{i + 1})
\end{aligned}
\end{equation}
The number of summands thus respect $ \mathrm{f}(2 k) = \mathrm{f}(k) $.

When $n$ is uneven: 
\begin{equation}
\begin{aligned}
    \left\{
    \begin{array}{ll}
        k & = 2^{w} + r \\
        n & = 2 k + 1 \\
        & = 2^{w + 1} + 2 r + 1
    \end{array}
    \right.
    \Rightarrow
    \left\{
    \begin{array}{ll}
        p & = n - 2^{w + 1} \\
        & = 2 r + 1 \\
        q & = 2^{w + 2} - n \\
        & = 2^{w + 1} - 2 r - 1
    \end{array}
    \right.
\end{aligned}
\end{equation}
Our induction follows:
\begin{equation}
    \mathrm{f}(n) = \mathrm{f}(2 r + 1) + \mathrm{f}(2^{w + 1} - 2 r - 1)
\end{equation}
which is also respected by Stern's sequence:
\begin{equation}
\begin{aligned}
    \mathrm{fusc}(2 k + 1) & = \mathrm{fusc}(2^{w} + r) + \mathrm{fusc}(2^{w} + r + 1) \\
    & = \mathrm{fusc}(r) + \mathrm{fusc}(2^{w} - r) \\
    & \quad + \mathrm{fusc}(r + 1) + \mathrm{fusc}(2^{w} - r - 1) \\
    & = \mathrm{fusc}(2 r + 1) + \mathrm{fusc}(2^{w + 1} - 2 r - 1)
\end{aligned}
\end{equation}




\section{Other Representations of the Graphs}
\begin{figure}[tb]
\begin{center}
\resizebox{\linewidth}{!}{\includegraphics{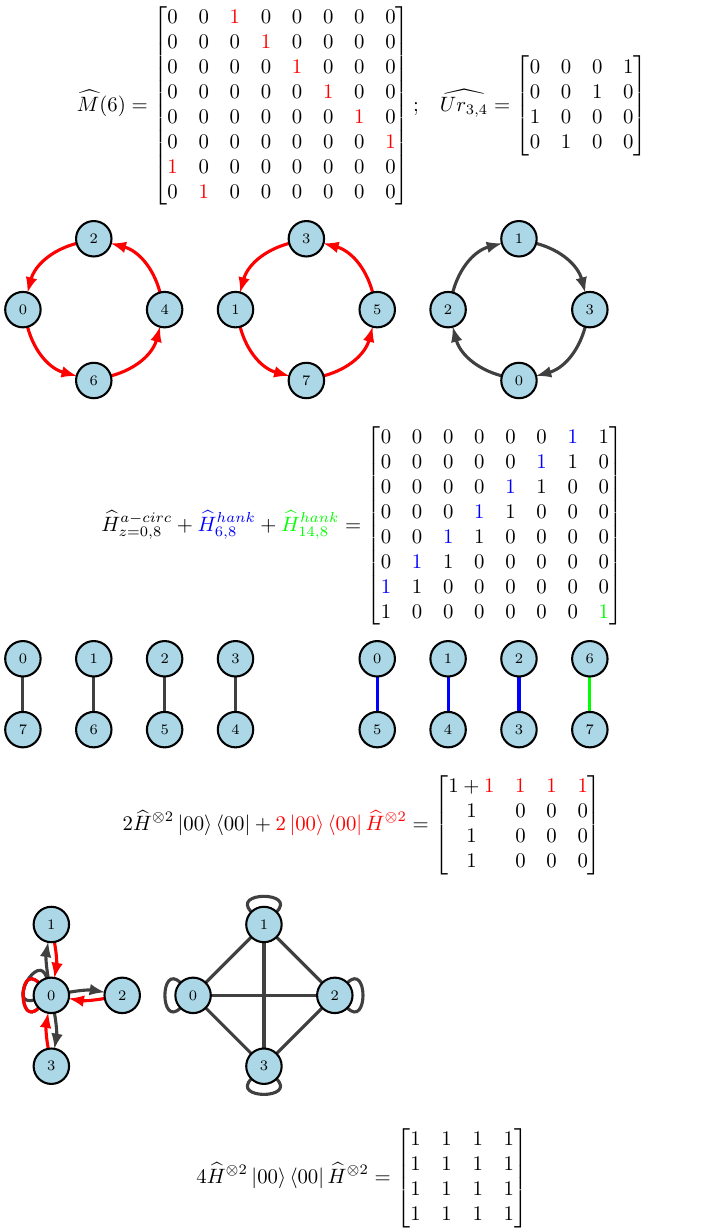}}
\end{center}
\caption{Other representations of the graphs that are discussed in the main article.}
\label{fig_graph_alternative}
\end{figure}

This appendix presents another visual representation of the graph presented in the main article \Cref{fig_graph_alternative}.
This alternative representation may help decompose the problem's graph into simpler graphs.

\section{Graph Construct as Cartesian Product of the Presented Graphs \label{appendix_graph_list}}
\begin{itemize}
    \item Ladder graph: $ \qop{H}_{m-1, m}^{Toep} \square \qop{X} $.
    \item Prism graph: $ \qop{H}_{m-1, m}^{circ} \square \qop{X} $.
    \item Book graph: $ \qop{S}_{m} \square \qop{X} $.
    \item Stacked prism graph: $ \qop{H}_{m-1, m}^{circ} \square \qop{H}_{n-1, n}^{Toep} $.
    \item Stacked book graph: $ \qop{S}_{m} \square \qop{H}_{n-1, n}^{Toep} $.
    \item Circulant graph: $ \qop{H}_{m-1, m}^{circ} \square \qop{H}_{n-1, n}^{circ} $ (not a cycle graph!) with $m$ and $n$ relatively prime (\textit{i.e.} $ \gcd(m, n) = 1 $).
    \item $m \times n$ rook graph:$ \qop{K}_{m} \square \qop{K}_{n} $.
    \item KP graph: $ \qop{K}_{m} \square \qop{H}_{n-1, n}^{Toep} $.
    \item KC graph: $ \qop{K}_{m} \square \qop{H}_{n-1, n}^{circ} $.
\end{itemize}

\end{document}